\RequirePackage{rotating}
\documentclass{egpubl}
 
\JournalSubmission    
\usepackage[T1]{fontenc}
\usepackage{dfadobe}  

\usepackage{cite}  
\BibtexOrBiblatex
\electronicVersion
\PrintedOrElectronic

\usepackage{egweblnk}

\usepackage{microtype}                 
\PassOptionsToPackage{warn}{textcomp}  
\usepackage{textcomp}                  
\usepackage{mathptmx}                  
\usepackage{times}                     

\usepackage{cite}                      
\usepackage{tabu}                      
\usepackage{booktabs}                  
\usepackage{tabularx}

\usepackage{multicol}
\usepackage{multirow}
\usepackage{siunitx}

\usepackage{xcolor}
\usepackage{colortbl}

\usepackage[percent]{overpic}

\definecolor{Gray}{rgb}{0.85,0.88,0.88}
\definecolor{LightGray}{rgb}{0.92,0.95,0.95}

\newcolumntype{a}{>{\columncolor{Gray}}c}
\newcolumntype{b}{>{\columncolor{LightGray}}c}

\makeatletter
\newcommand\notsotiny{\@setfontsize\notsotiny\@vipt\@viipt}
\makeatother

\newlength{\mylength}
\makeatletter
\newcommand{\mycfs}[1]{%
  \normalsize
  \@defaultunits\mylength=#1pt\relax\@nnil
  \edef\@tempa{{\strip@pt\mylength}}%
  \ifx\protect\@typeset@protect
     \edef\@currsize{\noexpand\mycfs\@tempa}
  \fi
  \mylength=1.2\mylength
  \edef\@tempa{\@tempa{\strip@pt\mylength}}%
  \expandafter\fontsize\@tempa
  \selectfont
}
\makeatother

\usepackage{array}
\newcolumntype{P}[1]{>{\centering\arraybackslash}p{#1}}

\usepackage{adjustbox}

\newcolumntype{R}[2]{%
    >{\adjustbox{angle=#1,lap=\width-(#2)}\bgroup}%
    l%
    <{\egroup}%
}
\newcommand*\rot{\multicolumn{1}{R{90}{1em}}}
\newcommand*\tB{$\bullet$}
\newcommand*\tb{}

\usepackage[inline]{enumitem}
\usepackage{rotating}

\title[Visualization on Large High-Resolution Displays]%
      {Interactive Visualization on Large High-Resolution Displays:\\A Survey}

\author[I. Belkacem \& N. M\'edoc \& S. Knudsen \& R. Dachselt \& M. Ghoniem]
{\parbox{\textwidth}{\centering 
        I. Belkacem$^{1}$
        and C. Tominski$^{2}$\orcid{0000-0001-7704-355X}
        and N. M\'edoc$^{1}$\orcid{0000-0002-7419-8748}
        and S. Knudsen$^{3}$\orcid{0000-0002-8306-1102}
        and R. Dachselt$^{4}$\orcid{0000-0002-2176-876X}
        and M. Ghoniem$^{1}$\orcid{0000-0001-6745-3651}
        }
        \\
{\parbox{\textwidth}{\centering $^1$Luxembourg Institute of Science and Technology, Luxembourg\\
         $^2$University of Rostock, Germany\\
         $^3$IT University of Copenhagen, Denmark\\
         $^4$Technische Universit\"at Dresden, Germany
       }
}
}

\begin{document}


\maketitle
\begin{abstract}
In the past few years, large high-resolution displays (LHRDs) have attracted considerable attention from researchers, industries, and application areas that increasingly rely on data-driven decision-making. An up-to-date survey on the use of LHRDs for interactive data visualization seems warranted to summarize how new solutions meet the characteristics and requirements of LHRDs and take advantage of their unique benefits.
In this survey, we start by defining LHRDs and  outlining the consequence of LHRD environments on interactive visualizations in terms of more pixels, space, users, and devices. Then, we review related literature along the four axes of visualization, interaction, evaluation studies, and applications. With these four axes, our survey provides a unique perspective and covers a broad range of aspects being relevant when developing interactive visual data analysis solutions for LHRDs.
We conclude this survey by reflecting on a number of opportunities for future research to help the community take up the still open challenges of interactive visualization on LHRDs. 


\end{abstract}  

\section{Introduction}

\maketitle

Screen real-estate is a key resource for visualization. So, visualization approaches naturally aim to use the available screen space efficiently~\cite{Tominski20IVDA}. Yet, conventional displays are limited in terms of how much information can be visualized at a time. To overcome this limit, researchers have investigated how physically larger displays with a larger number of pixels can be utilized for visualization purposes~\cite{andrews2011information}. Meanwhile, technological progress has made it cheaper and easier to build large high-resolution displays (LHRDs). 

Past research has shown unique benefits of LHRDs. The increased physical size and pixel resolution makes it possible to visualize massive data sets~\cite{ruddle2013leveraging}. The high resolution facilitates the representation of fine details alongside a general overview~\cite{isenberg2013hybrid}. The expanded interaction possibilities, such as mid-air gestures~\cite{matulic2018multiray}, gaze-based interaction~\cite{lander2015gazeprojector}, or physical navigation~\cite{lehmann2011physical}, offer whole new ways of working with data beyond what is possible in traditional settings. Moreover, the larger space in front of LHRDs allows people to collaboratively engage in data exploration and analysis activities~\cite{chokshi2014eplan}.

The visualization and human-computer interaction communities have actively studied LHRDs, as proven by a long history of publications. Previous surveys on LHRDs~\cite{ni2006survey,badillo2006literature,khan2011survey,andrews2011information,ardito2015interaction} already cover aspects of technology, visualization, and interaction, but separately so. In contrast, our survey integrates visualization, interaction, and also the stronger empirical foundations laid in the last decade. Based on a systematically extracted literature corpus, we provide an up-to-date view on LHRD research structured along the following questions:

\begin{itemize}[noitemsep]
    \item What are LHRDs and how are they characterized?
    \item How can data be visualized on LHRDs?
    \item How can one interact with data displayed on LHRDs?
    \item What empirical evidence exists for the usefulness of LHRDs?
    \item Where can LHRDs be applied successfully?
    \item What are open challenges for future research on LHRDs?
\end{itemize}

Answers to these questions are provided in the sections to come. In Section~\ref{sec:basics}, we characterize the properties of LHRDs and discuss their benefits in terms of visualization and interaction in more detail. We also identify the requirements and challenges that need to be addressed when pursuing visual data analysis on LHRDs.

At the core of this paper, Sections~\ref{sec:vis} and~\ref{sec:interact} review visualization and interaction approaches for LHRDs in detail. We describe how the unique benefits of LHRDs (e.g., larger size) are used and how the related challenges (e.g., interaction across larger distances) are addressed. Following the review of existing approaches, Section~\ref{sec:studies} sheds light on empirical studies that have investigated the use of LHRDs for interactive visualization. In Section~\ref{sec:applications}, we take a practitioner's view by illustrating several application scenarios taking advantage of LHRD visualizations. Finally, we identify and elaborate on research opportunities for future work in Section~\ref{sec:future}.

\paragraph*{Literature Corpus}

To find relevant literature, we conducted a systematic keyword search using general search queries (e.g., immersive analytics AND data visualization) and specific keywords (e.g., visualization wall, tiled display, wall-sized display) to ensure a broad bibliographic search. The queries were executed on six scientific databases, namely Springer, Wiley Online Library, ScienceDirect, ACM digital library, IEEE Xplore, EBSCO Host, which store most of the relevant computer science literature. 

We stored the query results in a spreadsheet, including information such as title, year, DOI, conference or journal, abstract, authors, keywords, search query, and database. At first, the spreadsheet contained 13,613 records, which were reduced to 8,632 after cleaning and deduplicating records. We further excluded records lacking an abstract, such as calls for participation, and keynotes. We screened the remaining records for relevance using the visual text mining tool Papyrus~\cite{medoc2016exploratory}. We spotted irrelevant topics (e.g., biological cell walls, brick walls) and consequently removed corresponding irrelevant venues, e.g., journals on biology or on construction. A total of 8,111 entries were retained after this cleaning step.

We then ranked the venues by decreasing popularity (from the CHI conference with 373 papers to 1,628 venues with only one paper). For venues having more than three papers, if at least one paper seemed relevant, all papers from that venue were retained, otherwise they were excluded. Venues with three or fewer records were dropped, unless the venue was clearly related to the VIS community and this survey.
Finally, one of the co-authors has screened the abstracts of the remaining $4,030$ papers published in $534$ distinct venues, to annotate them against exclusion/inclusion criteria fully described in \autoref{app:method}.
In the end, we kept $701$ papers.

This corpus of papers provides us with a broad view on interactive visualization on large high-resolution displays, but is still too large to be covered fully in a survey. Therefore, the individual co-authors used the Papyrus tool to further filter the literature and collect and classify prior work in ``shoeboxes'' for the different topics discussed in our survey (e.g., LHRD visualizations for different types of data or different layouts of LHRD views).
In addition to the formally collected literature, the co-authors also contributed further references based on their individual scientific background.
More details on the outlined methodology are given in \autoref{app:method}. 

\section{Large High-resolution Displays}
\label{sec:basics}

This section defines the key characteristics of LHRDs and examines what they imply with respect to interactive visualization.

\subsection{Definition}

When we refer to output devices as ``large high-resolution displays''~\cite{ni2006survey}, ``wall-sized displays''~\cite{lischke2015interaction}, or ``power walls''~\cite{rooney2013powerwall}, what do we actually mean?
In the literature, LHRDs are usually characterized based on two primary aspects: (i) the physical size and (ii) the pixel resolution~\cite{ni2006survey,andrews2011information,rooney2013powerwall}.
Accordingly, our definition is as follows:

\begin{quote}
{A large high-resolution display (LHRD) creates a coherent physical view space that is at least of the \textbf{size of the human body} and exhibits a \textbf{significantly higher resolution} than a conventional display.}
\end{quote}

This definition emphasizes the characteristics of the physical view space offered by display technology. To be considered an LHRD, the view space must be as big as or larger than a human. In other words, the view space covers or even extends beyond the human field of view, which also implies that parts of the view space may be beyond what humans can reach with their hands. Obviously, this rules out tablets, conventional desktop displays, and also some wall-mounted displays such as digital signage displays, whose view space is clearly below human scale. Creating an LHRD according to our definition often involves combining multiple displays. Tiled-display walls are a common example, where several smaller displays form a coherent larger one~\cite{beaudouin2012multisurface}. Other scenarios use several projectors to create an extended view space~\cite{schikore2000high}. In our survey, we focus on vertical LHRDs, which make the majority of existing devices.

In terms of pixel resolution, our definition resorts to a \emph{relative} statement in that it requires the overall resolution to be \emph{significantly higher} than for conventional displays (e.g., desktop monitor) of a given point in time. We refrain from defining \emph{absolute} pixel counts, as what was deemed a high resolution ten or twenty years ago is most likely no longer a high resolution now, and any number we state today would soon be outdated by the advance of technology. Accordingly, tiled-display walls or multi-projector displays have a higher resolution by design and qualify as LHRDs. Whereas physically larger interactive whiteboards as found in classrooms do not have a higher resolution and are hence not at the core of this survey.

\begin{table*}[t!]
	\mycfs{8}
	\centering
	\caption{Examples of LHRDs showing the evolution from early megapixel projectors setups to four-wall gigapixel display environments.}
	\setlength\tabcolsep{0.03cm}
	\begin{tabularx}{\textwidth}{@{} l X X X X X X @{}}
        &
        \includegraphics[width=\linewidth]{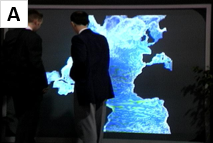} &
        \includegraphics[width=\linewidth]{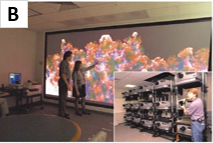} &
        \includegraphics[width=\linewidth]{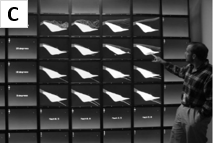} &
        \includegraphics[width=\linewidth]{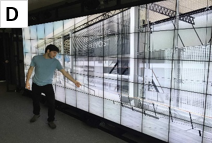} &
        \includegraphics[width=\linewidth]{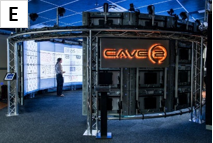} &
        \includegraphics[width=\linewidth]{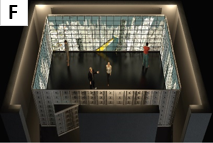} \\
        
		\hline
		\\[-6pt]
		\textbf{Name} & 
		Powerwall~\cite{PowerWall}    & 
		LLNL display~\cite{schikore2000high} & 
		Hyperwall~\cite{sandstrom2003hyperwall} &
		WILDER~\cite{beaudouin2012multisurface}       &
		CAVE2~\cite{febretti2013cave2}        &
		RealityDeck~\cite{papadopoulos2014reality}
		\\

		
		\textbf{Year} &
		1994 &
		2000 &
		2003 &
		2012 &
		2013 &
		2014 
		\\

		\textbf{Size} & 
		$1.82 \times 2.13$\SI{}{\metre} &
		$4.87 \times 2.44$ \SI{}{\metre} &
		\SI{5.10}{\metre\squared} &
		5.5 m $\times$ 1.8 m &
		6.7 \SI{}{\metre} diameter  &
		$10 \times 5.8 \times 3.35$ \SI{}{\metre}
		\\
		
		\textbf{Resolution} &  
		  8 Megapixels &
		 19 Megapixels &
		 64 Megapixels &
		131 Megapixels &
		 72 Megapixels &
		1.5 Gigapixels
		\\
		
		\textbf{Tiling} &  
        $2 \times 2$ &
        $5 \times 3$ &
        $7 \times 7$ &
        $8 \times 4$ &
        $18 \times 4$ &
        $16 \times 8 + 10 \times 8$
		\\

		\textbf{Technology\hspace{0.05cm}} &  
        Rear-projection  &
        Rear-projection &
        LCD &
        LCD &
        LCD &
        LCD
		\\

 		\hline

	\end{tabularx}
	\label{tab:lhrd_examples}
    \label{fig:lhrd_examples}
\end{table*}

In the scope of our definition, instances of LHRDs can have various physical properties. Table~\ref{tab:lhrd_examples} provides some examples with their key characteristics. 
Early LHRDs installations consisted of a matrix of projectors, e.g. at the University of Minnesota~\cite{PowerWall} or at the Lawrence Livermore National Laboratory (LLNL)~\cite{schikore2000high}.
Later, tiled screens were used to achieve higher resolutions, e.g., NASA's HyperWall~\cite{sandstrom2003hyperwall} or WILDER at University of Paris Saclay~\cite{beaudouin2012multisurface}. 
Some LHRDs offer an immersive experience such as CAVE2~\cite{febretti2013cave2}, a grid of screens arranged in a \ang{320} circle, or the RealityDeck~\cite{papadopoulos2014reality}, the first gigapixel LHRD with a rectangular arrangement around four walls.

\subsection{Implications for Visualization and Interaction}
\label{sec:proscons}

The properties of LHRDs as outlined above do have implications for visualization and interaction. In this section, we discuss briefly four key aspects: (1) more pixels, (2) more space, (3) more users, and (4) more devices. More details on visualization and interaction with LHRDs will be given in Sections~\ref{sec:vis} and~\ref{sec:interact}, respectively.

\subsubsection{More pixels}

In desktop environments, the small number of pixels can lead to situations where a full understanding of the data is difficult to gain, because not all relevant information can be visualized at once. To circumvent these problems, one can look at different visualizations successively generated through interaction, which may be tedious. 

LHRDs can show much more data at a time, which limits the need to generate different visual representations interactively. The greater number of pixels of LHRDs makes it possible to visualize in parallel more data items, data variables, data facets, and data scales.

When there are more pixels, a visualization can show more data before over-plotting occurs. As over-plotting is reduced, details can be discerned more easily, while an overview is maintained at no extra costs. The larger screen real-estate of LHRDs brings the opportunity to lay out multiple complementary views each emphasizing a particular data facet. Multiple views are also useful when analyzing data at different scales. For comparative analyses, it is easily possible to show even larger parts of the data side by side on an LHRD. All these possibilities have led to proven gains in terms of quantity and quality of insights and sense-making~\cite{reda2015effects,andrews2010space}, and the ability to notice more details~\cite{rajabiyazdi2015understanding}.
 
\paragraph*{Consequences}

The availability of more pixels means that visualization designers have to think about how to spend the pixels most efficiently. Simply showing more is not enough. There are still the limits of human perception and cognition. Care must be taken not to overwhelm users with too much information. The visualizations and the data items within them must be properly organized so that the relevant information can be grasped easily. Also the technical limits must be considered. Visualizing more data also requires more computational power. This calls for efficient graphics hardware as well as data structures and algorithms. Standard visualization libraries often do not address the specific technical requirements of LHRDs. Instead, special-purpose solutions need to be employed~\cite{Mei20DataV}.

\subsubsection{More space}
\label{subsubsec:more-space}

The larger physical size of LHRDs affects how visualizations are consumed. Users turn out to be more engaged in their tasks, more relaxed about their ability to perform their work~\cite{bi2009comparing}, and more effective at certain tasks as display size increases~\cite{lischke2015using}. Unlike for regular desktop environments, users of LHRDs also have a wide field of view which facilitates the perception of more data at a time.

There is also more space in front of LHRDs. Hence, users are free to move physically to see different regions of the display, or to move closer to get more details or farther out to get an overview. This ability is known as \emph{physical navigation}, as opposed to \emph{virtual navigation} via pan and zoom. Physical navigation has the advantage that users can explore data naturally as they would inspect objects in the real world, and this can lead to increased performance~\cite{ball2007move}.

Visualization research has leveraged physical navigation as an interaction modality. So-called proxemic interaction exploits user distance, orientation, movement, and location, e.g., to change the visual encoding, to cast dynamic queries, or to adapt the interaction fidelity~\cite{lehmann2011physical,jakobsen2013information,dostal2014spidereyes, kister2015bodylenses}. One can use physical navigation to create hybrid multi-scale visualizations, where detailed information is blended into the overview such that the user can read off more details as they get closer or embrace the overview as they step back~\cite{isenberg2013hybrid}. 
FatFonts also use a symbolic visualization to reveal details or delivers an overview, depending on the distance between display and viewer~\cite{Nacenta12FatFonts}.

Studies found that physical navigation in front of LHRDs is intuitive, boosts user performance, and is preferred to virtual navigation~\cite{ball2007move,ball2007realizing,liu2014effects,jakobsen2015moving}. Physical navigation also plays a greater role than peripheral vision for certain tasks~\cite{ball2008effects,jansen2019effects}. In a way, movement unlocks the use of different cognitive resources, which were tied, e.g., to a better use of spatial memory~\cite{radle2013effect}.

\paragraph*{Consequences}

More space has pros and cons for visualization. Due to the large physical size, some parts of the visualization may be far away or be presented at extreme angles. This can affect the perception of visual variables~\cite{bezerianos2012perception,bezerianos2013perceptual}, an issue that might be mitigated by curved rather than flat LHRDs~\cite{prouzeau2016visual}. Yet, as users may move in front of the display, viewing distances and angles vary, which complicates the design of visual representations that can always be interpreted accurately. Also, user interface (UI) elements may be out of reach. This calls for visualizations that dynamically adapt to the user's position, e.g., by relocating selected data items or UI elements to the vicinity of the user~\cite{badam2016supporting}. Finally, moving in front of LHRDs implies higher physical costs when using a visualization~\cite{Lam08InteractionCosts}. The visualization designer needs to carefully balance the advantages of physical navigation with its costs.

\subsubsection{More users}

Analyzing complex data coming from various sources may require multiple experts to collaboratively examine the data from several complementary standpoints. Regular desktop environments can hardly fit more than one user in front of the display and do not support simultaneous multi-user interaction. Hence, discussions between users and the confrontation of ideas are much hampered.

The larger space offered by LHRDs can fit many users in front of the same visualization. LHRDs can accommodate a continuum of collaboration styles, ranging from parallel work, to discussions, to close collaboration~\cite{liu2014effects,radloff2015supporting,wallace2016creating, langner2018multiple}. LHRDs also facilitate working with public and personal private information spaces~\cite{von2014sleed}. Beyond collocated collaboration, mixed presence collaboration extends the scope of application of LHRDs~\cite{avellino2015accuracy,avellino2017camray,marrinan2017mixed}. 

Collaboration between users leads to a shared awareness of the data and improves task completion. Teamwork in front of LHRDs was found comfortable and helpful for answering questions~\cite{langner2018multiple}. Paired analytics studies show that task completion is faster and more accurate with pairs than with individuals~\cite{prouzeau2016evaluating}.

\paragraph*{Consequences}

Collaboration is a demanding concern when designing visualization and interaction for LHRDs. The visual representation must allow multiple users to study the data, while occlusion of the data by users close to the LHRD must be mitigated~\cite{Radloff11SmartViews}. The interface must be designed so that public and private working spaces, or territories (cf. \cite{Scott04Territoriality}),  are available on demand. Suitable interaction modalities must be integrated. In a shared working environment, individual users must be identifiable to correctly attribute an interaction or to provide individual UI tools, and conflicting actions must be handled~\cite{von2016youtouch}. Besides these non-trivial technical challenges, collaboration adds totally new social aspects to visual data analysis. For one, the communication and discussion among users must be considered. Measures are needed to ease team work and motivate all team members to work on a shared goal~\cite{jakobsen2016negotiating}. Also, negative feelings due to competitive conditions should be avoided~\cite{mayer2018pac}. All these technical and social aspects explain why collaborative data analysis on LHRDs is still a challenging research endeavor.

\subsubsection{More devices}

LHRDs are often built as a combination of several devices, including displays for output and sensors for input. More output devices lead to the benefits in terms of display space and high-resolution described earlier. LHRDs can also be used with additional smaller displays to complement the large display space~\cite{kister2017grasp,horak2018david}. In smart rooms, for example, people can bring their own devices, which are then integrated seamlessly into the environment~\cite{radloff2015supporting}.

More input devices lead to a rich environment for multi-modal interaction where commands can be issued in many ways. Touch is but one interaction modality found on LHRDs. Proxemic interaction is another modality, which requires sensors tracking user position in front of LHRDs. Cameras can capture user movements to ease interaction via mid-air gestures~\cite{matulic2018multiray}. Spoken commands can also be useful when analyzing data on large displays~\cite{Srinivasan21MultiModal}.

\paragraph*{Consequences}

The fact that multiple devices need to work in concert in a common infrastructure is a key technical challenge for visualization on LHRDs. On the output side, the first question is how to render complex visualizations at interactive frame rates on many displays at once. This requires tailored and potentially distributed rendering solutions~\cite{chung2013survey}. Also, visual discontinuities across devices must be compensated to faithfully depict data~\cite{li2000building,hereld2000introduction}. Indeed, display bezels act as occluders of information and as obstacles for interaction~\cite{ebert2009tiled++,de2012looking}. Analog to distributing visualization content to several displays, user actions must be collected from various input devices~\cite{tse2003rapidly,pietriga2011rapid,Radloff12InteractionManagement}. Besides handling output and input, an overall coordination of the hardware and software environment is also needed~\cite{marrinan2014sage2, rittenbruch2015supporting}. Still, many of these technical issues are hard to deal with. While research prototypes exist for individual concerns~\cite{Mei20DataV}, so far there is no standard approach for turnkey visualization on LHRDs.

To recap, this section defined what LHRDs are and characterized them along four perspectives: more pixels, more space, more devices, and more users. While this section looked briefly at LHRDs in the context of interactive visualization, the next Sections~\ref{sec:vis} and \ref{sec:interact} provide more details on visualization on and interaction with LHRDs.
\section{Visualization on LHRDs}
\label{sec:vis}

LHRDs bring about many opportunities and challenges for the design of visualization systems.
While standard visualization software may often be run as-is on LHRDs, prior visualization and interaction designs are likely to be unsuited and may frustrate the user.
More relevant visualization designs are needed to harness the increased resources and mitigate the potential challenges brought by LHRDs.

In this section, we draw the landscape of visualization research on LHRDs.
Based on our review, we first characterize in Section~\ref{sec:layout} four fundamental approaches that might be applied to exploit the larger display space offered by LHRDs. In Section~\ref{sec:visualization-techniques}, we then survey existing visualization approaches across six different types of data.
The gist of this section consists in answering the question, whether a visual data representation has been explicitly designed for LHRDs or not.
If so, how has the existing work accomplished this?

\subsection{Utilizing the Larger Display Size}
\label{sec:layout}

The main benefit of LHRDs is their larger screen real-estate, which allows more data to be shown at once. We identify four main approaches for the use of the larger display size: 

\begin{description}
    \item[Single view:] The available display space is used exclusively to show a single high-resolution visualization.
    \item[Small multiples:] The display space is split into multiple smaller views showing different subsets of the data using the same visual encoding (e.g., multiple choropleth maps for different years).
    \item[Multiple views:] The display space is partitioned into multiple views each showing data with a different visual encoding (e.g., choropleth map, parallel coordinates, node-link diagram showing the different aspects of a multi-faceted graph).
    \item[Distributed views on multiple displays:] The large display is used in concert with one or more auxiliary displays with the option of showing different parts of the data in different ways.
\end{description}

In the next sections we expose how these four approaches relate specifically to LHRDs, and how they leverage the larger pixel counts, the space around the displays, and how they support collaboration.

\subsubsection{Single view: all in one view}

The \emph{single view} approach aims to use the display space to show a single data visualization. The huge high-resolution visuals can help users in many fields interpret large and complex data. Networks, genomics data, geographic information systems, and gigapixels images used in earth sciences and medicine are just a few examples. 
The larger amount of information displayed at once has been proved to boost task completion and user satisfaction~\cite{reda2015effects}.
In a way, data visualization has turned into collaborative large-scale data visualization, with collaboration as one of the main challenges~\cite{Thomas05IlluminatingPath}.

\begin{figure}[t]
\includegraphics[width=\linewidth]{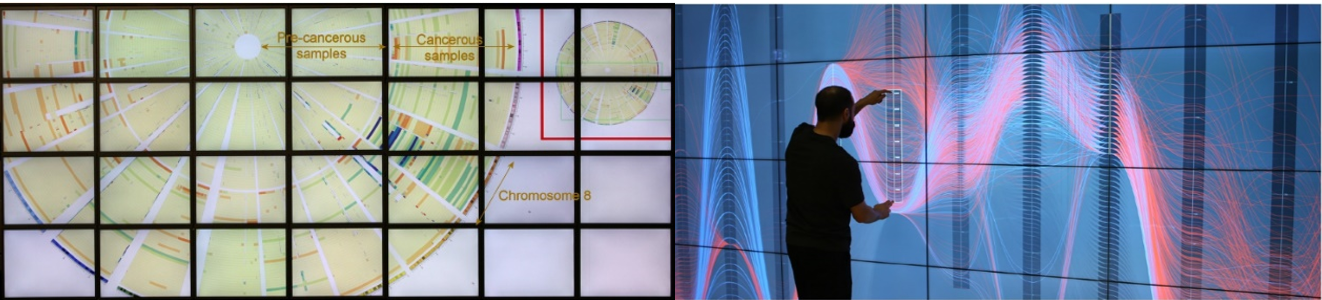}
\caption{Single high-resolution views. (left) A circular heatmap for comparative genomics analyses~\cite{ruddle2013leveraging}; (right) parallel coordinates plot of a movie dataset~\cite{Reibert2020-PACM}.}
\label{fig:Vis_HighRes}
\end{figure}

A common example of high resolution single view is navigating maps. Users perform better at higher pixel counts when utilizing maps on large displays compared to normal monitors with restricted physical size and pixel count~\cite{ball2005evaluating}. \autoref{fig:Vis_HighRes} (left) from~\cite{ruddle2013leveraging} shows a Circos heat map about chromosomes. Unlike LHRD, on a desktop display the user had to pan 12 times just to see the whole graphic, and so was unlikely to notice the same features. Another example in \autoref{fig:Vis_HighRes} (right) from~\cite{Reibert2020-PACM} shows that the enhanced display size and resolution of LHRDs can help parallel coordinates visualizations since they can handle more dimensions and data elements. In the healthcare domain, pathologists examine tiny slices of tissue under a microscope, usually at a magnification of 25–400, to identify illnesses such as cancer. This produces images at resolutions over 1 Gigapixel (e.g., $32,000 \times 32,000$). Displaying such data on LHRDs turns out to be a most promising option to support pathologists in their work~\cite{goodyer2009using,ruddle2016design}.

\begin{figure}[t]
\includegraphics[width=\linewidth]{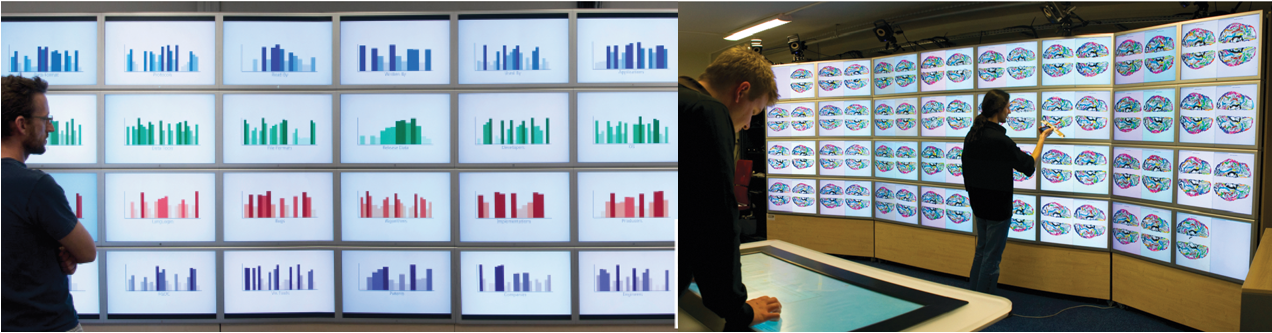}
\caption{Small multiples on an LHRD. (left) A grid of 32 histograms~\cite{bezerianos2012perception}; (right) a grid of 64 3D brain scans~\cite{gjerlufsen2011shared}.}
\label{fig:Vis_SmallMultiples}
\end{figure}

\subsubsection{Small multiples: same encoding for different data}

\emph{Small multiples} are well-established across both desktop and large display visualizations. They are ``shrunken, high-density graphics based on a large data matrix''~\cite{Tufte83VisualDisplay}. They are a collection of comparable miniature graphs or charts showing different subsets of the data with the same visual encoding (see \autoref{fig:Vis_SmallMultiples} left). They are ``often narrative in content, showing shifts in relationship between variables as the index variable changes''~\cite{Tufte83VisualDisplay}. The advantage of using small multiples on LHRDs is that the individual views need not be small. In fact, they can be as large as or even larger than single-view visualizations on regular desktop displays. As shown in \autoref{fig:Vis_SmallMultiples} (left), small multiples are often aligned with the natural tiling of the individual displays in an LHRD.

Figure \ref{fig:Vis_SmallMultiples} (right) shows an example of small multiples of 3D brain scans~\cite{gjerlufsen2011shared}. Neurologists compared and classified a collection of scans on an LHRD, moving the pictures to ease comparisons.  
With LHRD, the charts take the space they need allowing multi-user scenarios. For many problems, small multiples are the best design solution. It has been shown that for navigating time intervals on LHRD, small multiples are faster than other visualizations~\cite{lee2020effectiveness}.

\subsubsection{Multiple views: rich encoding of complex data}

\begin{figure}[t]
\includegraphics[width=\linewidth]{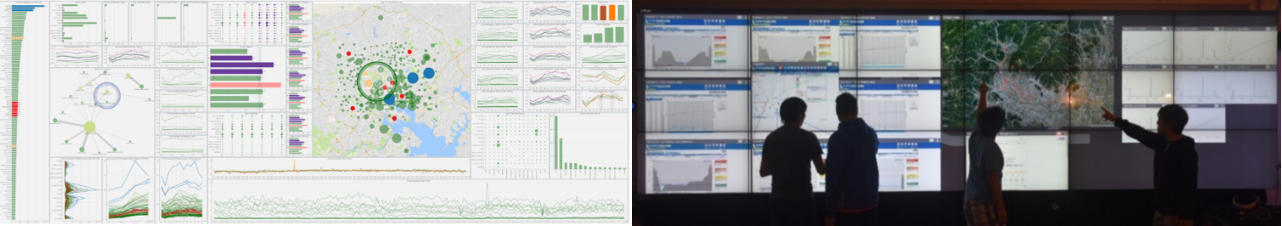}
\caption{Multiple views on LHRDs. (left) 47 coordinated views of multivariate crime data~\cite{langner2018multiple}; (right) plots and maps showing sensor data~\cite{kobayashi2018sage}.}
\label{fig:Vis_MCV}
\end{figure}

When analyzing complex phenomena users often need to work with big, multi-dimensional, heterogeneous, and dynamic data coming from different sources in parallel . Relying on a single view can lead to complex visualizations that are hard to understand~\cite{Tominski21FlexibleVA}. In contrast, \emph{multiple views} enable users to explore data from different angles~\cite{Roberts07CMV, langner2018multiple}. Each view may show a different aspect of the data or use a different visual encoding that complements other views (see \autoref{fig:Vis_MCV} left). An LHRD can fit more views, and more high resolution visualizations in each view, than its replica in a conventional desktop setting. This adds new visual design requirements, e.g., a parsimonious use of colors~\cite{langner2018multiple}. 

Multiple views are usually coordinated, in which case we speak of multiple coordinated views (MCV). Coordination means that interactions performed in one view are automatically propagated to other views. Selecting and highlighting data elements is one common coordination, for which several sorts of view relationships might exist (e.g., overview+detail, difference views).

Applying the concept of multiple views on LHRDs allows each individual view to benefit from the larger display space and thus to alleviate some of the issues with multiple views on regular displays concerning screen space, computer performance and user perception~\cite{wang2000guidelines}. 
LHRD can also support practical applications that require simultaneous consideration of static data and continuous data streams like dashboards with multiple users. This complexity cannot be handled with regular displays that impede user understanding.

\subsubsection{Distributed views: visualization on multiple displays}

Finally, LHRDs can be part of multi-display environments in which users can perform their tasks using \emph{distributed views} on various devices, including tabletops, laptops, tablets, or smartphones. Typically, the LHRD is used as a public information radiator of relevant information for all users~\cite{chokshi2014eplan}. A semi-public tabletop may then serve as a cooperation area, and small private screens may show role-specific information. For such distributed views, some information can be complementary or even duplicated across devices.

\begin{figure}[t]
\includegraphics[width=\linewidth]{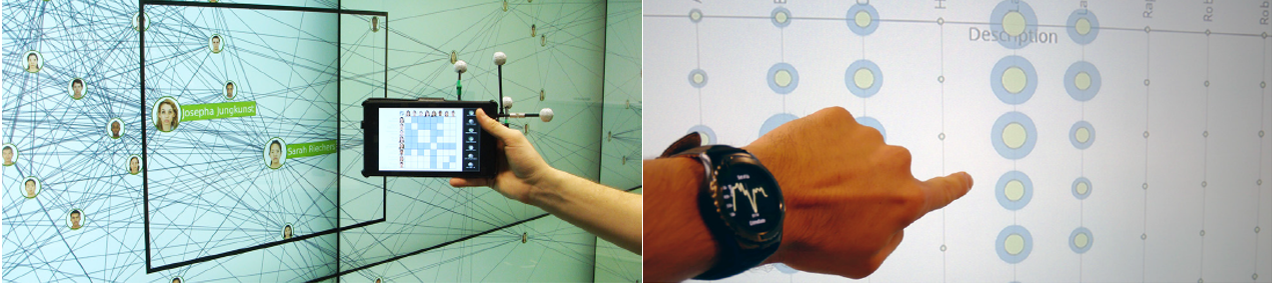}
\caption{Examples of LHRDs in multi-display environments. (left) A mobile device showing an alternative representation of a sub-graph selected on the LHRD~\cite{kister2017grasp}; (right) interaction to transfer selected data from LHRD views to the user's smartwatch~\cite{horak2018david}.}
\label{fig:Vis_DUI}
\end{figure}

An example is to use the LHRD as an overview of the entire information space and personal devices to display details of user-selected objects (see \autoref{fig:Vis_DUI} left). This may include close-up excerpts, adjusted level of detail, or alternative representations of the selected data~\cite{kister2017grasp}. Role-specific data can be displayed on personal devices in scenarios where different users with different expertise need to analyze different parts of the data~\cite{prouzeau2018awareness}.

Utilizing public views and personal, private views avoids disruptions and clutter on the LHRD~\cite{von2014sleed,von2015using}. Private views give users ``space to think'' and avoid interference with other users, for example, when applying view changes or filters. In such scenarios, interactions affect only the visualization on the private smartphone~\cite{von2014sleed}, tablet~\cite{sollich2016exploring}, or smartwatch~\cite{horak2018david}, with the option to apply changes to the LHRD on demand (see \autoref{fig:Vis_DUI} right). Interestingly, different user behaviors can be observed during collaboration in multi-display environments~\cite{kister2017grasp}. Some users distribute their work evenly on the available devices, frequently switching between the LHRD and the mobile display, while others focus very much on the mobile device.

\subsubsection{Discussion}

The use of single high-resolution views and multiple views pose unique challenges regarding perception from different angles and interaction reachability. Users may be unable to access, visually and interactively, data items at the top or the far end of an LHRD. Also, the amount of information that can be displayed at once may be staggering. That said, automatically deciding which views are to be shown on which region is as important as techniques for supporting the user to focus on relevant data. Several points can be considered:
\begin{enumerate*}[label=\arabic*)]
    \item the process of creating views and inserting them into the environment;
    \item the creation of groups of views; 
    \item defining a layout for the views. 
\end{enumerate*}
While predefined layouts can be used, some visualization tasks may require dynamic layout changes, e.g., by adding views or moving views as users move in front of the display~\cite{radloff2015supporting}. Regarding distributed views, a coordination mechanism across devices is important to support the access to the involved views~\cite{langner2018multiple}.

One advantage is that users may naturally zoom in and out by assuming close or far positions to the LHRD and pan by physically moving left and right. Walking in front of the LHRD is an apparent benefit which supports exploration of different views and support sense-making. We will further discuss interaction in \autoref{sec:interaction}.

In sum, the \emph{single view} is the most homogeneous setup; all of the display space is used to show one high-resolution visualization.
Less homogeneous is the \emph{small multiples} setup where the space is tiled evenly to display multiple visualizations of the same nature, e.g., for comparative analysis.
The \emph{multiple views} setup tiles a set of views in a way that gives some views more space than others.
They support complex analytical workflows where different visualizations cover different aspects of the data.
Lastly, the \emph{distributed views} setup lays out the visualizations on adjunct display devices, besides the wall display itself.
In \autoref{app:tables} \autoref{tab:visualization_tab} we show how selected works use the space of LHRDs.
The table is not meant to be exhaustive, but rather exemplifies the design space of visualization on LHRDs. Next, we continue with visualization on LHRDs for different types of data.

\subsection{LHRD Visualization for Different Data Types}
\label{sec:visualization-techniques}

Inspired by prior visualization taxonomies~\cite{card1999readings, munzner2014visualization}, this section is organized based on data types.
We cover the visualization of geographical data, spatial data, temporal data, multidimensional data, network and tree data, and text and document data. We offer an overview of selected relevant techniques in \autoref{app:tables}, \autoref{tab:visualization_tab}. Due to space constraints, we only describe a subset of the listed techniques below.

\subsubsection{Geographical data visualization}

First, we reflect on the use of LHRDs for map-based visualizations of data associated with locations relative to Earth, so-called  geo-spatial or geographical data. 
Maps of such data are used in many application areas. 
Owing to their familiarity, maps have often been used in HCI research to assess benefits of LHRDs, for various tasks from object lookup, and route tracing~\cite{ball2005evaluating} to more elaborate insight finding~\cite{reda2015effects} and collaborative exploration~\cite{su_visually_2018}.

\begin{figure}[t]
\includegraphics[width=\linewidth]{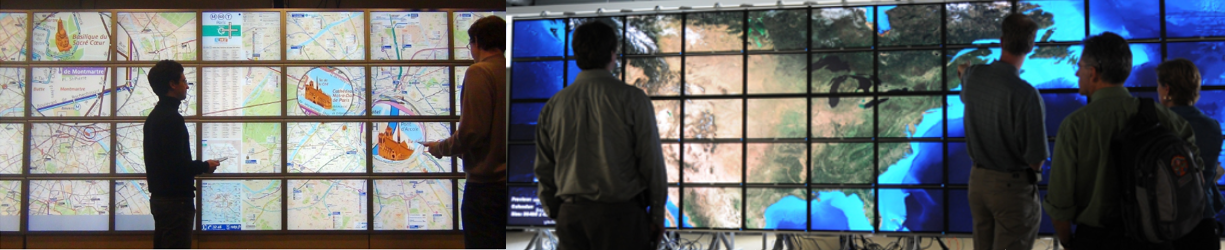}
\caption{Geographical data visualization. (left) Users control a magnification lens on a map using mobile devices~\cite{chapuis2014smarties}; (right) users collaborate on exploring a large map~\cite{weibel2010hiperpaper}.}
\label{fig:vis_maps}
\end{figure}

Map visualizations benefit directly from the higher resolution since a bigger part of the map could be visualized at a time and interactively at multiple scales~\cite{ball2005evaluating, ball2007move,jakobsen2013information}. 
Many types of visual overlays are also used with maps, like traffic data~\cite{prouzeau2016towards}, social media messages~\cite{onorati2015walltweet} and network visualization~\cite{booker2007high}, which is also true on LHRDs.
The higher resolution results in less visual occlusion when showing details on demand~\cite{ball2005evaluating} or when enriching the map with information in focus+context designs~\cite{ball2008effects,chapuis2014smarties} (see \autoref{fig:vis_maps} left), which both can be also achieved with additional personal displays \cite{von2014sleed}.

Besides visual overlays, maps are also juxtaposed with other visualizations in MCV settings~\cite{langner2018multiple}. The larger space can fit several maps at once, corresponding to different regions under scrutiny by multiple users, and at multiple scales, e.g., in emergency response scenarios. Yet new design challenges arise in terms of spatial layout and spatial grouping of views~\cite{langner2018multiple}.
Also, when the LHRD is not touch enabled, users may prefer paper maps~\cite{chokshi2014eplan}.

For map visualizations on LHRDs, the larger space in front of the display triggers spontaneous physical navigation (see \autoref{fig:vis_maps} right), which is preferred by users to virtual navigation and boosts user performance~\cite{ball2007move, ball2007realizing}. 
The reason is that physical navigation unlocks the use of other types of `embodied resources', e.g., spatial memory~\cite{ball2008effects}.
Also a body scale display supports natural user interactions, e.g., using body shadows to reach for tools and store data~\cite{shoemaker2010whole, kister2015bodylenses}. 
Seeing oneself and the surrounding world also boosts performance for navigation tasks on LHRDs, e.g., compared to high-resolution head-mounted displays~\cite{ball2007move}.

LHRDs have also been used for collaborative map-based work to support information sharing, e.g., for situational awareness~\cite{chokshi2014eplan}.
A natural extension of map visualizations on LHRD consists in showing transient information about the activity of other users on a shared display in the form of awareness bars on the sides of the map or hulls showing past and present focus of users in a selected region of the map~\cite{prouzeau2018awareness}.
Persistent floor displays showing the footsteps of other users were also explored~\cite{prouzeau2018awareness}.

While suitable for displaying and navigating map data,
we seem to lack visualization techniques designed specifically for the perceptual challenges of LHRDs.
Such work include multiscale typographic visualizations like FatFonts~\cite{Nacenta12FatFonts} and hybrid-image visualizations~\cite{isenberg2013hybrid}.
Both exploit the higher resolution to provide in the same picture several levels of legibility of information according to the viewing distance, for maps and other types of data visualizations.

\begin{figure}[t]
\includegraphics[width=\linewidth]{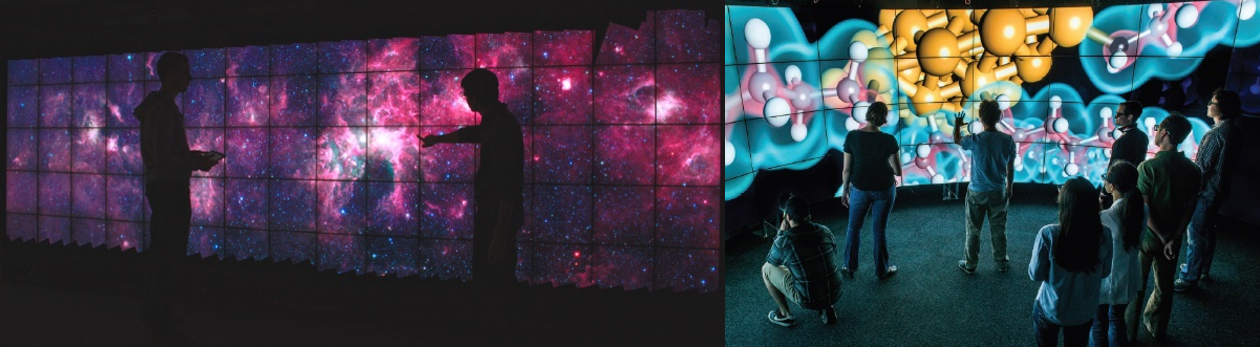}
\caption{Spatial data visualization. Experts explore (left) a set of high-dynamic-range astronomical images~\cite{pietriga2016exploratory}; (right) simulations of 3D volumes of electron density around atoms~\cite{reda2013visualizing}.}
\label{fig:Vis_Physical}
\end{figure}

\subsubsection{Spatial data visualization}

Other types of spatial data describe objects having a geometry, but whose geographic position is irrelevant to the analysis at hand.
This includes computer-aided design (CAD) objects, 2D and 3D medical imagery, as well as field and volume data.
The visualization of such non-geographical spatial data serves many areas such as life sciences~\cite{goodyer2009using}, cosmology~\cite{hanula2015cavern,pietriga2016exploratory}, engineering~\cite{chung2013developing} and education~\cite{johnson2006geowall,jagodic2011enabling}.

With more pixels and more space, spatial data may be presented more effectively on LHRD to improve the understanding of complex spatial relationships.
The higher resolution of LHRD supports the visual analysis of a single high resolution graphical representation of large data sets captured by modern instruments, notably gigapixel images in medicine~\cite{goodyer2009using,treanor2009virtual,ruddle2016design} and astronomy~\cite{pietriga2016exploratory} (see \autoref{fig:Vis_Physical} left). 
Like for maps, a larger portion of the image can be shown at once compared to desktop monitors and can be augmented with other data modalities such as text labels and navigation reference points.
Visual thumbnails are used in overview+detail interfaces to support navigation~\cite{goodyer2009using}. 
Multifocal fisheye distortion was also used to magnify multiple regions of interest in large astronomy data sets~\cite{pietriga2016exploratory}.
Moreover, the higher resolution affords the use of small multiples at a larger scale, e.g., for the comparative analysis of a deck of brain scans taken at different angles or time points~\cite{gjerlufsen2011shared}, or large trajectory data sets~\cite{reda2013visualizing}.
Similarly, dozens of coordinated views can be used to support the analysis of multiple complementary data sets at once, e.g., for genome-sequencing~\cite{reda2013visualizing}.

LHRDs have been combined with other devices, e.g., stereoscopic glasses to visualize 3D objects, e.g., ball-and-stick molecular models~\cite{reda2013visualizing} (see \autoref{fig:Vis_Physical} right) or terrain models~\cite{johnson2006geowall,chung2013developing}.
Handheld devices are also used as secondary displays in complement to LHRDs, e.g., to visualize the details of a selected subset of data whereas the wall display shows the context~\cite{sollich2016exploring, horak2018david}.
Yet having to switch attention between the wall display and a handheld personal display may hamper user performance~\cite{tan2003effects}.
More research is needed to elicit workflows and tasks where a secondary personal display is helpful~\cite{horak2018david}.

The use of LHRDs to visualize spatial data speeds up decision-making, conveys more insights from the data, and encourages users to reflect on meaning and behavior~\cite{treanor2009virtual,reda2015effects,ruddle2016design,langner2018multiple}.
It makes for a more pleasurable, engaging, and educational experience~\cite{reda2015effects}.
The interaction capabilities of LHRDs may also provide a more immersive experience for spatial data~\cite{reda2013visualizing}.

\subsubsection{Temporal data visualization} 
\label{sec:time-vis}

Like space, time is a key aspect to understand many phenomena. 
This makes temporal data analysis and visualization important in many domains~\cite{Aigner11TimeVis}.
Temporal data can grow very large.
The longer the time period being considered, the more time points in the data, and the larger the data set.
Since temporal data often exceed the display capacity of regular screens,
LHRDs seem promising for the analysis of temporal data. 
Oddly our corpus on LHRDs contains only a few temporal data visualization papers (Supplemental Table~2).

For example, the hybrid-image visualization technique~\cite{isenberg2013hybrid} was used to visualize 22 years of temperature data on an LHRD. 
From afar, the user sees barcharts of the average monthly temperature in a small multiples layout.
Up close, the user perceives individual line charts of daily temperatures, without any changes in the graphic representation due to user movements.
The computational cost of this technique makes it only suited for static representations.

\begin{figure}[t]
\includegraphics[width=\linewidth]{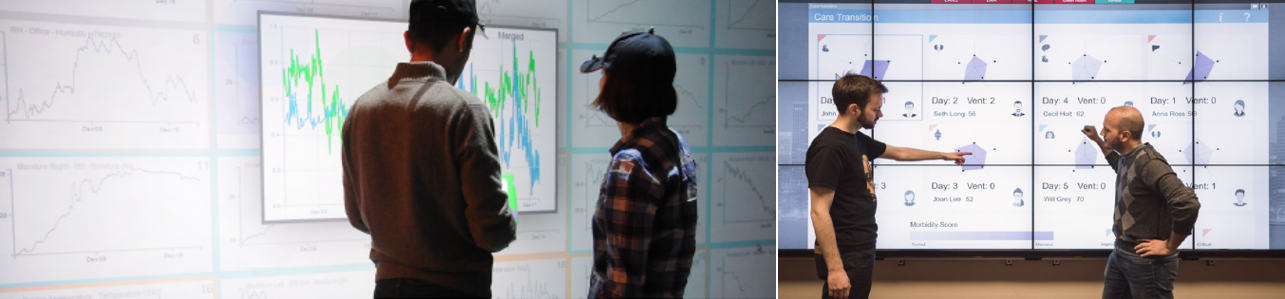}
\caption{Temporal data visualization. Collaborative visual analysis of (left) multi-sensor time-series in a building~\cite{badam2016supporting}; (right) a summary overview of patient data over time~\cite{thomas2017echo}.}
\label{fig:Vis_Temporal}
\end{figure}

More examples of temporal visualizations on LHRDs concern patient data from intensive care units~\cite{thomas2017echo} (see \autoref{fig:Vis_Temporal} right), spatio-temporal crime data~\cite{reda2015effects,langner2018multiple}, and various types of sensor data from buildings~\cite{badam2016supporting} (see \autoref{fig:Vis_Temporal} left). 
All these examples use MCV where the temporal data visualization is one view among other views showing other aspects of the data, e.g., a geographic map or a density plot.
In this sense, the large space of LHRDs is mostly used to enable the analysis of further data aspects in relation to time.
So far, there has been little research dedicated to utilizing the advantages of LHRDs for temporal data specifically.

\begin{figure}[t]
\includegraphics[width=\linewidth]{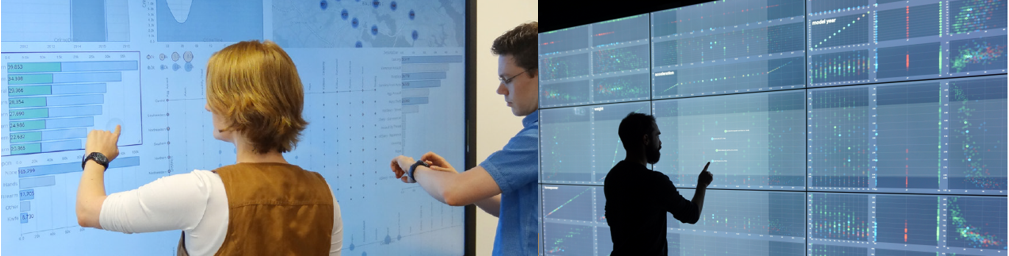}
\caption{Multidimensional data visualization. (left) Crime data visualized in multiple views~\cite{horak2018david}; (right) SPLOM representing car data~\cite{Riehmann2020-CGF}.}
\label{fig:Vis_Multidimensional}
\end{figure}

\subsubsection{Multidimensional data visualization}

High-dimensional data has motivated much work in data science and data visualization.
From a visual mapping perspective, existing techniques can be grouped in four main categories: axis-based, glyph-based, pixel-based and, hierarchy-based visualizations~\cite{Tominski20IVDA}.
Popular techniques are the axes-based parallel-coordinates plots (PCP) and scatterplot matrices (SPLOM)~\cite{liu2016visualizing}.

Building on the increased resolution and space, SPLOM have been used on LHRDs in a single-view setting~\cite{Riehmann2020-CGF}. Either more pixels and space are given to individual scatterplots to reduce visual clutter, or to fit more scatterplots at once (see \autoref{fig:Vis_Multidimensional} right).
The display space can also be used to enhance the points in a scatterplot by nesting glyphs encoding details legible from up close~\cite{chegini2017interaction}.
SPLOMs may also be one view in a multiple-views setting~\cite{chegini2017interaction, langner2018multiple, tsandilas2015sketchsliders, horak2018david} (see \autoref{fig:Vis_Multidimensional} left).

Used in full-screen mode on LHRD, PCP have been extended with specific user interactions to overcome reachability and clutter problems for single-user analyses~\cite{Reibert2020-PACM}. 
For PCP, possible complex problem-solving scenarios include for different users to work on separate axes, or on the same axis, which covers brushing and linking, virtual navigation and view configuration tasks. 

Prior work has focused on interaction and collaboration rather than visual encoding and perceptual challenges of LHRDs. Collaborative data analysis offers many research avenues.
Various types of multivariate visualizations raise various challenges and opportunities in terms of collaboration style, interaction and visual encoding.

\subsubsection{Network and tree visualization}

\begin{figure}[t]
\includegraphics[width=\linewidth]{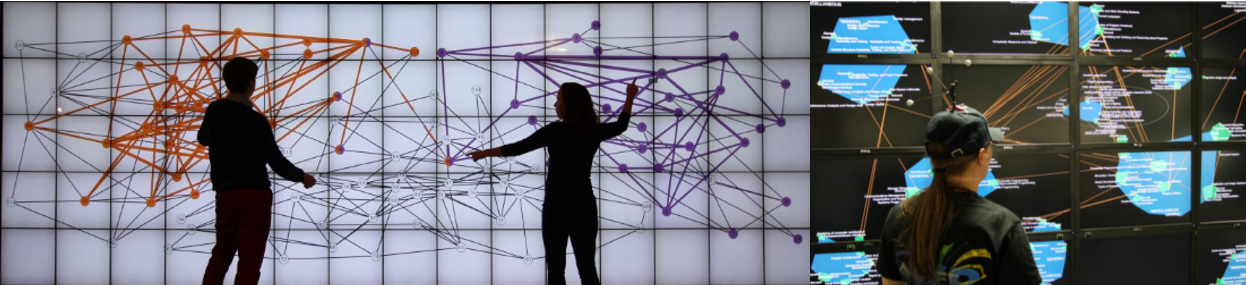}
\caption{Network and tree visualization. (left) Collaborative network exploration on an LHRD~\cite{prouzeau2016evaluating}; (right) exploring a hierarchical network using lenses~\cite{lehmann2011physical}.}
\label{fig:Vis_Network}
\end{figure}

Graphs are used to model relationships between entities and analyze intricate data patterns.
They are well studied in visual analytics research~\cite{von2011visual}.
They are often large and dense, which results in much visual clutter on standard displays and hampers sense-making.
The larger resolution of LHRD promises to reduce node and edge density in the visualization~\cite{prouzeau2016evaluating} (see \autoref{fig:Vis_Network} left).

The hybrid-image technique already mentioned earlier can also be applied to display multiscale graph visualizations~\cite{isenberg2013hybrid}. To this end, high-level edges and labels are rendered in a way to be visible from a distance but become less visible close to the display. Viewers can then see group-level relationships from a distance, while relationships between individual nodes are seen up close. 

Graph visualization on LHRDs often comes with additional tools, e.g., magic lenses and focus+context to support the data exploration~\cite{kister2015bodylenses}. For example, access to different levels of detail can be based on a handheld device~\cite{kister2017grasp} or the user's physical position in front of an LHRD~\cite{lehmann2011physical} (see \autoref{fig:Vis_Network} right). Dedicated selection techniques may further support multiple users to explore different parts of the graph visualization all at once~\cite{prouzeau2016evaluating}. 

\subsubsection{Visualization of texts and documents}

\begin{figure}[t]
\includegraphics[width=\linewidth]{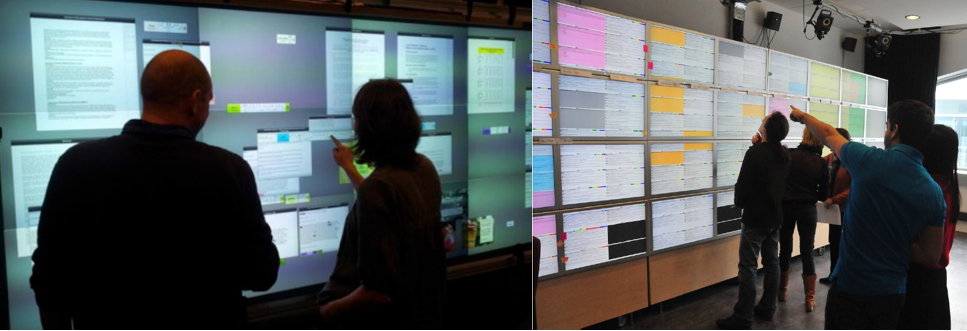}
\caption{Text and document visualization. (left) Exploring a news corpus~\cite{jakobsen2014up}; (right) Conference chairs classifying and fine-tuning a large number of sessions and research papers~\cite{liu2014leveraging}.}
\label{fig:Vis_Text}
\end{figure}

Text data can be of interest per se, e.g., news articles~\cite{jakobsen2014up} (\autoref{fig:Vis_Text} left) or scientific papers~\cite{liu2014leveraging} (\autoref{fig:Vis_Text} right).
Also, text labels help to interpret other data visualizations.
While the legibility of text is critical~\cite{vinot_legible_2012}, occlusions can quickly occur~\cite{onorati2015walltweet, ellsworth_interactive_2017}.
Especially on desktop displays, the visualization designer is often torn between showing the semantic structure of the corpus~\cite{liu_topicpanorama_2014} or the detailed textual content~\cite{don_discovering_2007}.

LHRDs improve the sense-making of text data by exploiting spatial memory to offload working memory and by using spatial layout to encode semantic relationships~\cite{andrews2010space}. 
With document-centric approaches where multiple views embed textual contents, the larger space helps to get an overview~\cite{jansen2019effects} and to lay out documents freely~\cite{andrews2010space,kirshenbaum2019side}, which reduces memory load. 
It also boosts collaboration by better separating responsibility~\cite{bradel2013large}. 
But in shared areas users need to discuss more and some group dynamics may be frustrating~\cite{birnholtz2007exploratory}.

With visualization-centric approaches, e.g. Jigsaw~\cite{stasko_jigsaw_2008}, users move less in the space and work more independently~\cite{bradel2013large}. Yet, owing to the externalization of semantic relationships, memorization and computation tasks are replaced by more efficient perception tasks~\cite{andrews2010space,geymayer_how_2017}.
Hybrid approaches can both reveal useful patterns and contextualize them with textual content~\cite{fiaux_bixplorer_2013,chokshi2014eplan,chung2014visporter}.
The level of detail can be adjusted using physical navigation~\cite{ andrews_analysts_2012}, which also increases user movements~\cite{jakobsen2012proximity} and triggers spatial memory~\cite{jakobsen2014up}. Semantic interaction (term highlighting, spatial grouping of documents, annotations) may improve the visual analytics workflow \cite{endert_semantic_2012-1}.

Exploring large and/or streaming text data requires an effective combination of LHRD-based visualization, automated analytic processing, and collaboration. For example, LHRDs were used to assist users in exploring social media postings~\cite{su_visually_2018, fernando_towards_2020}, news articles~\cite{andrews2010space} or open government data~\cite{kukimoto_open_2014}.

Labels are small text snippets that help users interpret the data.
The larger space of LHRD helps to provide a tight spatial coupling of labels and data~\cite{polys2007effects}.
To maintain the readability of labels from different distances, label font size can be varied depending on the user's position and viewing direction~\cite{lehmann2011physical}.
Label readability can also be supported through hybrid images, which provide different levels of details depending on the viewing distance~\cite{isenberg2013hybrid}.

Still little attention was given to factors influencing user performance in reading and text exploration tasks~\cite{nutsi_readability_2016, iyer_text_2017, kirshenbaum2019side}.
Future LHRD research could study factors such as font size~\cite{lischke2015using, liu2014effects}, direction and speed of moving text, and information density~\cite{vinot_legible_2012}. 
Understanding the pros and cons of different text visualization approaches on LHRDs for different tasks and text collections of different size remains an open challenge.

In this section, we presented four key approaches to exploit the larger space of LHRDs and reviewed concrete visualizations for different types of data and how they leverage LHRDs.
From our review, there seems to be few visualization techniques whose visual encoding was designed specifically for LHRDs, with hybrid images~\cite{isenberg2013hybrid} being one example. While popular visualization techniques on the desktop have been simply transferred to LHRD environments, approaches like \cite{Reibert2020-PACM, Riehmann2020-CGF, kister2017grasp} are careful adaptations for LHRD environments.

This also includes novel ways of interacting with LHRD visualizations, where traditional single user desktop interaction is mostly not suitable. Instead, diverse interaction techniques and approaches for multi-modal and collaborative exploratory data analysis are focused on, which will be the topic of the next section.

\section{Interaction}
\label{sec:interact}
\label{sec:interaction}

Mouse and keyboard, the main input devices for desktop visualization, are often neither available nor suitable or useful on LHRDs.
Consequently, researchers have explored and proposed a range of possibilities for interaction that address the specific requirements of human-scale interaction with and in front of LHRDs: 

\begin{itemize}
    \item Both close-range and distant-range interaction must be considered since users often switch between close display proximity to look at details and distant display proximity to gain an overview. 
    
    \item Remote manipulation techniques are needed besides direct manipulation as some areas of the display may be unreachable. Frequent actions must be possible from several display positions at once.
    
    \item Parallel input channels, multi-user support, or collaborative interaction techniques are often needed since LHRDs support parallel work or collaboration on either shared or individual display areas.

    \item Data spaces can be navigated by moving around in front of a display (physical navigation) or by using body movement as input to adapt the visualization (proxemic interaction), since LHRDs are often installed in rooms with space in front of them.
    
    \item The increased number of users, the wide availability of personal mobile devices, and the need for distant interaction suggest the usage of LHRD as one central part of multi-display environments. 
    
\end{itemize}

To provide a common vocabulary for understanding what and how the different input and interaction techniques support analytic needs, we start by introducing the seven general interaction tasks by Yi et~al.~\cite{yi2007toward} and put them in context with LHRDs:

\begin{description}
    \item[Select] \emph{-- Mark something as interesting}. 
    Marking one or more items in a visualization, to differentiate selected and unselected items. 
    
    LHRDs lead to stronger needs to mark and keep track of items of interest during immersive analysis sessions, but challenges arise, for example, in that selections can be performed by multiple concurrent users. 
    Selection can be supported by a range of input technologies, which raises new challenges, such as how to design precise and efficient selection techniques, also from the distance.
    
    \item[Explore] \emph{-- Show me something else}.
    Altering the viewpoint, thus changing which data items are visible or not, e.g., using panning. 
    
    LHRDs can reduce the need for virtual panning, which can be replaced by moving from one part of a display to another. But, if information spaces are larger than the LHRD, virtual panning may not be work well as changing the viewport across an LHRD can lead to motion sickness. Designers may instead rely on multiple views that allow smaller areas to be explored virtually.
    
    
    \item[Reconfigure] \emph{-- Show me a different arrangement}. 
    Changing the position of visual items, like data marks, structural marks, and views, e.g., through manual spatial organization or layout techniques.
    
    With LHRDs, visualization items can be distributed across large areas. 
    Yet, rearranging data marks across large areas of a display can be disorienting and must be done carefully. 
    Also, LHRDs can show many views at once.
    Doing so amplifies the need for managing, organizing, and navigating these views. 
    In response, both manual and automatic view layout approaches might be useful.
    For example, organizing views manually allows people to use ``space to think''~\cite{andrews2010space}, and automatic layout techniques may ease the display of many views on LHRDs.
    Yet, automatic layout approaches must factor in physical distances between views, e.g., in how a layout might support view comparison or not. 
    
    \item[Encode] \emph{-- Show me a different representation}. 
    Changing the visual encoding of a data set or changing the visualization technique entirely, thus altering how data is shown. 
    Yi et al. distinguish between changes to how data attributes are mapped to visual variables (such color, size, and shape) and changing the visualization technique (e.g., a pie chart instead of a histogram).
        
    Changing the visual representation across an LHRD can be a powerful way to explore and understand data sets. It might yet cause confusion between collaborators being affected by this global change. Thus, local changes or personal views might work better in some situations.
    Changing the visualization technique on an LHRD might be less useful. Instead, the ample space can fit different complementary views side by side.
    
    \item[Abstract/Elaborate] \emph{-- Show me more or less detail}. 
    Modifying the level of abstraction, thus altering how much data is shown for different data points.

    LHRDs make it especially possible to show many details for each data point. 
    Yet, interaction techniques such as geometric zoom or drill-down interaction can still be relevant. In addition, semantic zoom techniques taking into account the varying distances of users to the display might be particularly useful for LHRDs.
    
    \item[Filter] \emph{-- Show me something conditionally}. 
    Changing criteria for which data are shown, thus changing which data items are visible or not, e.g., based on whether data items are in a range or not. 
    
    LHRDs offer extra space and resolution to show many data points at once. While positive in many cases, this can also be staggering and distracting from the essential. Thus, it can often be more relevant to filter out data items. Ideally, in multi-user settings, such ability should also be offered to individual users independently. 
    
    \item[Connect] \emph{-- Show me related items}. 
    Choosing to show associations and relationships between already shown data elements or to show additional data items relevant to a specified item.
    
    Again, LHRDs excel at showing multiple linked views. 
    Brushing, for example, may be used to highlight selected data items in one view across other views.
    Yet, existing techniques for MCV may not always scale well to LHRDs.
    For example, highlighting data items across an LHRD may be confusing for coworkers or raise perceptual challenges due to the sheer display size.
    Also, using lines to connect data marks across an LHRD amplifies risks of occlusion, while it might be hard to see both ends of the line. 
    
\end{description}

Clearly, the seven interaction tasks relate to visualizations on LHRDs and pose special challenges. Yi et al. further admit that ``other interaction techniques in InfoVis systems certainly exist''. For LHRDs, one may consider, e.g., interacting with between-view meta visualizations~\cite{knudsen2016view} to help users to make sense of views and their relationships. Visualization provenance~\cite{ragan2016characterizing} can also support users in keeping track of analysis goals, progress and insights.

Next, we discuss in detail \textit{how} the outlined tasks can be achieved either close to or directly on the vertical display (section \ref{subsec:on-surface-interaction}), from afar (section \ref{subsec:distant-interaction}), or by using the space in front of the display (section \ref{subsec:spatial-front-LHRD}). 
We further consider three forms of \emph{multiplicities} in the context of LHRDs:
\textit{multiple} displays, e.g., in the form of mobile devices
(section \ref{subsec:MDE-interaction}),
\textit{multiple} modalities, e.g., the mix of touch and mid-air gestures
(section \ref{subsec:multi-modal-interaction}),
and \textit{multiple} users, like custom interaction techniques for collaborative work 
(section \ref{subsec:multi-user-interaction}).

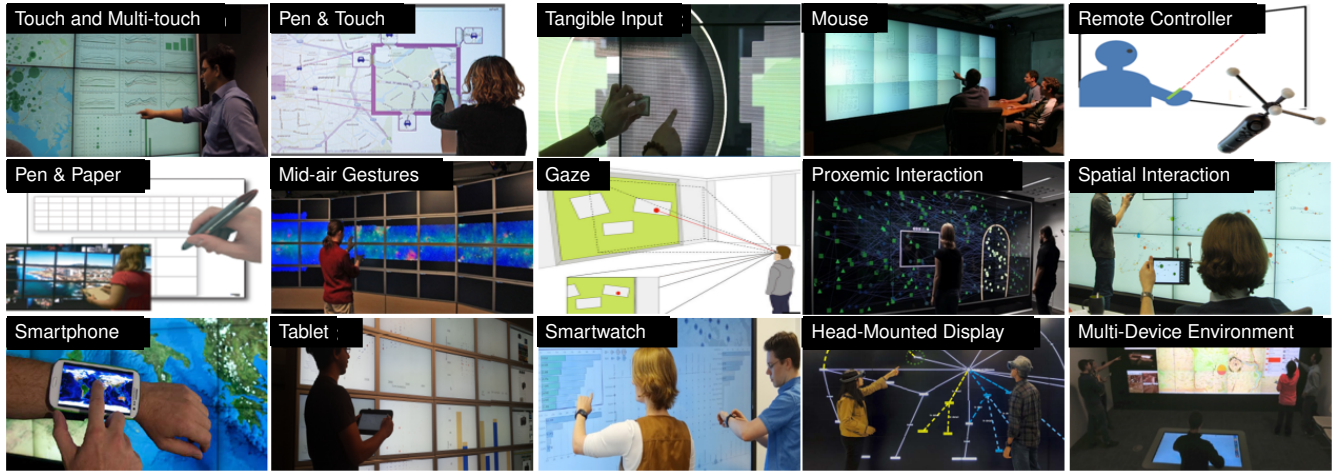
\begin{figure*}[t!]
\centering
\begin{overpic}[width=\linewidth,tics=10]{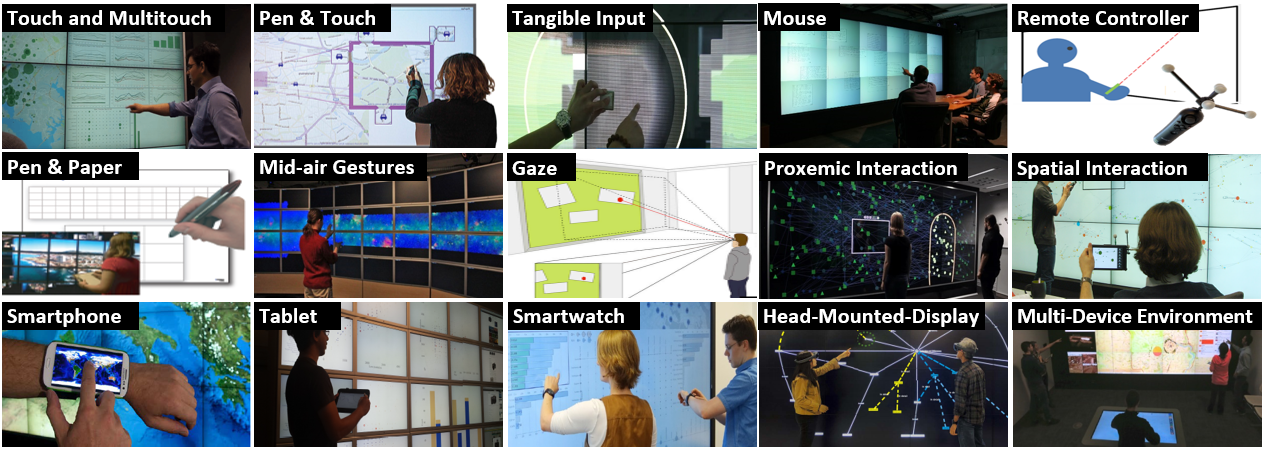}
 \put (0.2,33.9) {\textsf{\colorbox{black}{\textcolor{white}{\small{Touch and Multi-touch~~~}}}}}
 \put (20,33.9) {\textsf{\colorbox{black}{\textcolor{white}{\small{Pen \& Touch~~}}}}}
 \put (40,33.9) {\textsf{\colorbox{black}{\textcolor{white}{\small{Tangible Input~~}}}}}
 \put (60,33.9) {\textsf{\colorbox{black}{\textcolor{white}{\small{Mouse}}}}}
 \put (80,33.9) {\textsf{\colorbox{black}{\textcolor{white}{\small{Remote Controller~~~}}}}}

 \put (0.2,22.2) {\textsf{\colorbox{black}{\textcolor{white}{\small{Pen \& Paper~~~}}}}}
 \put (20,22.2) {\textsf{\colorbox{black}{\textcolor{white}{\small{Mid-air Gestures~~~}}}}}
 \put (40,22.2) {\textsf{\colorbox{black}{\textcolor{white}{\small{Gaze}}}}}
 \put (60,22.2) {\textsf{\colorbox{black}{\textcolor{white}{\small{Proxemic Interaction~~~~}}}}}
 \put (80,22.2) {\textsf{\colorbox{black}{\textcolor{white}{\small{Spatial Interaction~~~}}}}}

 \put (0.2,10.4) {\textsf{\colorbox{black}{\textcolor{white}{\small{Smartphone~~~}}}}}
 \put (20,10.4) {\textsf{\colorbox{black}{\textcolor{white}{\small{Tablet}}}}}
 \put (40,10.4) {\textsf{\colorbox{black}{\textcolor{white}{\small{Smartwatch~~~}}}}}
 \put (60,10.4) {\textsf{\colorbox{black}{\textcolor{white}{\small{Head-Mounted Display~~~~}}}}}
 \put (80,10.4) {\textsf{\colorbox{black}{\textcolor{white}{\small{Multi-Device Environment~~~~}}}}}
\end{overpic}
\caption{Interaction: touch and multi-touch~\cite{langner2018multiple}, pen and touch~\cite{ion2013canyon}, tangible input~\cite{Zadow2016miners}, mouse~\cite{birnholtz2007exploratory}, remote controller~\cite{jota2010comparison}, pen and paper~\cite{weibel2010hiperpaper}, mid-air gestures~\cite{nancel2011mid}, gaze~\cite{lander2015gazeprojector}, proxemic~\cite{kister2015bodylenses}, spatial~\cite{kister2017grasp}, smartphone~\cite{von2014sleed}, tablet~\cite{tsandilas2015sketchsliders}, smartwatch~\cite{horak2018david}, head-mounted-display~\cite{sun2019collaborative}, multi-device environment~\cite{chokshi2014eplan}.}
\label{fig:interaction}
\end{figure*}

\subsection{On-surface Interaction}
\label{subsec:on-surface-interaction}
Unlike early LHRDs, many recent installations allow direct on-surface interaction, thus uniting input and output space.
Modern built-in or added-on technologies, e.g., infrared frames, allow sensing touch, multi-touch, and pen input, rather rarely also tangibles.

\subsubsection{Touch and multi-touch}

Direct touch interaction requires standing close to the display.
Several users can touch the LHRD at once~\cite{jakobsen2014up}, which leads to better collaboration, user enjoyment, precise interaction, awareness of others and user satisfaction~\cite{jakobsen2016negotiating,heidrich2011interacting}. ``People's first action when seeing a 'bright shiny display' is to touch it.''~\cite{langner2018multiple}

Yet, standing near the display reduces the field of view.
Users must move back and forth to \emph{abstract} and get an overview or \emph{elaborate} and see details up close.
They must also move to \emph{explore} distant areas of the display, which may be tiring~\cite{jakobsen2015should}. 
The top of the display may be unreachable for users to interact (e.g., for tasks like \emph{select}, \emph{filter} or \emph{annotate}~\cite{prouzeau2016evaluating}).
Collisions and arm entanglements may occur as several users interact with a small shared display, which leads them to either negotiate for space or avoid reaching for areas near their coworkers~\cite{jakobsen2016negotiating}. 
In Jakobsen et al.'s work~\cite{jakobsen2014up}, users \emph{explore} a news corpus such that touching an article highlights all related articles. Users can search for and \emph{filter} relevant articles to make hypotheses and gather
evidence and \emph{connect} the articles. Touching and dragging with a finger lets the user \emph{reconfigure} the layout of the articles, while  a lengthy touch allows the user to annotate them. 
Drag and drop is commonly used to \emph{reconfigure} items spatially, which is unpleasant for the fingers, especially when the starting point and the target are far apart. One solution is to create temporary proxies of distant targets and bring them near the cursor for quick and easy reach~\cite{baudisch2003drag,collomb2005improving,bezerianos2005vacuum,doeweling2010drop}.

For visual comparison tasks on LHRD, users can use techniques that replicate a small region of a visualization and bring it close to them~\cite{tan2004wincuts,bezerianos2007using} or display a widget to interact with a remote area using different controls~\cite{khan2004remote}.
Langner et~al.~\cite{langner2018multiple} use popular direct touch gestures to interact with multiple coordinated views by touch. For example, a single tap is used to \emph{select} a data item, a drag to \emph{reconfigure} the item, and a pinch and two-finger drag for \emph{abstract/elaborate} navigation.
Many interactive visualizations featuring touch input seem to be designed for handheld devices and tabletops (for the latter, a transition to LHRDs is often possible). 

Among the few touch approaches designed for visualization on LHRDs, Reibert et~al.~\cite{Reibert2020-PACM} and Riehmann et~al.~\cite{Riehmann2020-CGF} describe techniques for interacting with parallel coordinates plots and scatter plots, respectively. This includes short-contact and multi-touch gestures such as fling on selection to \emph{filter}, swipe along an axis to \emph{select} a range, tap to \emph{elaborate} and drill down into details, two-finger fling to \emph{explore}, or pinch on the plot to \emph{reconfigure} the layout.
These gestures address reachability issues and aim to avoid uncomfortable prolonged touch gestures on LHRD surfaces.

One must note that true multi-touch is under-explored (except for the omnipresent pinch-to-zoom gesture), although the space on a large vertical display would easily allow it as opposed to smaller mobile devices. The aforementioned dedicated multi-touch interactions for parallel coordinates and scatter plots~\cite{Reibert2020-PACM,Riehmann2020-CGF} or the multi-touch gestures for fluently interacting with information visualization lenses~\cite{Kister:2016:MultiLens} are exceptions to this.
More work is still needed to develop comprehensive multi-touch interaction vocabularies for data visualization on LHRDs.

\subsubsection{Pen input}
Direct touch interaction is also possible with pens, though they serve other purposes.
While fingers can be preferred for direct manipulation, pens are well suited to \emph{annotate}, including sketching visualizations and taking handwritten notes~\cite{guimbretiere2001fluid,Walny2012-TVCG, lee2015sketchinsight,Romat:2019:Th-Inking}. They also allow new ways of \emph{selecting} data marks (e.g., by circling or crossing) or \emph{filtering} data (e.g., by striking through data items or drawing upper/lower limits directly).
Pen interaction does not suffer from the fat finger problem of touch, 
and pens do not leave traces or fingerprints on the display. Some pen technologies can also identify multiple users working simultaneously on the LHRD. Pen gestures, variance in pressure, tilt angle, or pen-holding grip \cite{Drini:2018:pen-postures} are other interesting degrees of freedom, though they have not been specifically used for LHRD visualization yet.

Pens can for example be used to \emph{select} an item on a map or to \emph{explore} other parts of the map~\cite{ion2013canyon}.
Handwriting recognition and word recognition, e.g., variable or function names, can support \emph{filtering} tasks. For example, Matulic et~al.~\cite{Matulic:2014} use pen-drawn sketches and other map annotations for intuitive and effective spatial querying of geographical data with user-specified scopes. 
Also, recognizing particular symbols, like an arrow, is used to move or clone a chart (i.e., \emph{reconfigure} the layout) or a set of points to \emph{change the encoding} from a barchart to a scatterplot~\cite{lee2015sketchinsight}.
Annotating parts of a visualization with a pen (and touch) for sensemaking activities, to externalize thoughts \cite{Kim:2019:Inking} and to support data analytics actions by using ink strokes \cite{Romat:2019:Th-Inking} are other natural uses of pen input for data visualizations on large vertical displays.

\subsubsection{Tangible input}
Tangibles have rarely been used to interact with LHRDs.
As most LHRDs are vertical displays, tangibles cannot lie on them like on horizontal tabletops. 
Unlike touch displays, tangibles allow to distinguish users by associating them with their own tangible marker~\cite{Zadow2016miners}.
The need for users to keep holding the tangibles is yet a limitation.
Tangibles based on magnets~\cite{leitner2011geckos} or vacuum self-adhesion~\cite{hennecke2012vertibles} are easier to use on vertical surfaces. 
WallTokens~\cite{courtoux2021walltokens} leverage 3D printing and inexpensive materials, like springs and suction cups, to produce a tangible exhibiting a multi-touch pattern when in contact with a tactile surface.
It can be left in place on a vertical LHRD surface without falling.
Tangibles can act as controllers for data \emph{selection}, \emph{exploration}, and \emph{filtering}. Users may \emph{reconfigure} the space and organize data items by drag and drop. Associating each person with a distinct tangible is an effective means of recognizing users and supporting interaction based on their roles in the team~\cite{courtoux2021walltokens}.

\subsection{Distant Interaction}
\label{subsec:distant-interaction}
The larger physical size of LHRDs raises issues like the reachability of parts of a visualization, or that users have to move away from the display to get an overview, as described in section \ref{subsubsec:more-space}. Users need techniques to interact remotely, e.g., using traditional input devices, dedicated remote controllers, mid-air gestures, and gaze input.

\subsubsection{Mouse and keyboard}
Early visualization research on LHRDs focused on input modalities designed for desktops, like mouse and keyboard interaction~\cite{andrews2010space}. 
With a mouse users can engage with an overview visualization and carry out their work from a distance while being seated~\cite{birnholtz2007exploratory}. 
They can also move freely in front of the display using a wireless mouse~\cite{jakobsen2016negotiating}.
Mouse interaction is faster and more accurate than touch~\cite{jakobsen2016negotiating}.
Yet, clutching may be a problem when having to travel long distances on an LHRD, as is the visibility of the small mouse pointer.
For WIMP (Windows, Icons, Menus, Pointer) interfaces, pointing is a key operation to \emph{select} items and perform other interaction tasks on the selected items on an interactive visualization (e.g., \emph{reconfigure}, \emph{encode}, \emph{filter}, etc).
When there are more pixels, distant target acquisition by moving a cursor over a long distance or when the target is small can be time consuming according to Fitt’s law~\cite{soukoreff2004towards}. Acquisition time can be shortened on LHRDs by creating multiple cursors and moving the closest cursor to the target~\cite{kobayashi2008ninja} or by increasing the target size~\cite{mcguffin2002acquisition,ramos2007pointing}, but the latter has not yet been explored for LHRD.

A mouse pointer can be used across multiple surfaces in a collaborative environment even when the environment is driven by several computers, like in PointRight~\cite{johanson2002pointright}.
Mouse pointers support various interaction tasks as in desktop environments.
Either a single mouse pointer is shared between different users, or each user has a distinct mouse pointer. A shared mouse pointer promotes more discussion, but may frustrate the users who do not control the mouse. In contrast, multiple mouse pointers allow parallel work, but harms discussion quality~\cite{birnholtz2007exploratory}.
The use of multiple mouse pointers can also lead to item selection conflicts, more than direct touch, since mouse pointers run unrestrained across the display.

Keyboards are mainly used to enter text, e.g., annotations. Text legibility is limited from a distance with a physical keyboard. Yet, up close text entry on an LHRD may be tedious with a virtual keyboard.

\subsubsection{Remote controllers}
\label{subsec:remote-controllers}
Handheld devices can be used as remote controllers to carry out diverse tasks from varying distances. 
When used for eyes-free interaction, they allow users to focus on the material shown on the LHRD. 
In early days, laser pointers served as an intuitive pointing device that used raycasting to support natural interaction to \emph{select items} and perform all other tasks based on selection ~\cite{davis2002lumipoint,konig2007position}. 
Free of clutching, raycasting is more practical given the long distance a cursor may travel on an LHRD~\cite{langner2016content, kister2017grasp}. Different colors and tags can be used to support multi-user contexts.
As technology evolved, relative or absolute pointing can be carried out on the touch surface of handheld devices~\cite{nancel2013high, von2014sleed} or with Vive controllers~\cite{zhang2017combining}, custom designed prototypes~\cite{baudisch2006soap, Klamka2015:Elasticcon} or flysticks fitted with reflective markers~\cite{jota2009comparative,jota2010comparison}.

Smartphones and smartwatches are widespread and can extend the Bring-Your-Own-Device (BYOD) paradigm to include user interaction. By touching the device screen without looking at it, users can perform tap, hold, or swipe actions~\cite{langner2018multiple,horak2018david, langner2016content}. Horak et al.~\cite{horak2018david} cover a wide range of interaction tasks using smartwatch interaction such as swiping or rotating a physical control of the smartwatch. This interaction allows users to apply \emph{filters} to data items, delete them by wiping them, and \emph{explore} the data and \emph{elaborate} on additional details on demand.

Eyes-free interaction is very relevant to interaction with LHRDs since the visual attention is already occupied with the large display. Techniques with haptic feedback~\cite{nancel2013high} (e.g., mouse wheel), tangible interaction~\cite{jansen2012tangible} (e.g., tangible sliders) or other physical constraints~\cite{Klamka2015:Elasticcon} (e.g., cord-based elastic interaction) can promote eyes-free interaction by making it simple, rapid and unwavering.
 
Besides being used as eyes-free touch interaction devices, the spatial position and orientation of mobile devices, like smartphones and tablets, can also be exploited to interact with LHRDs~\cite{langner2016content,langner2018towards}.
For example, the device position can be used to \emph{select} the region of interest on the LHRD~\cite{kister2017grasp}, to \emph{explore} using panning, or to \emph{abstract and elaborate} by zooming, e.g., tablet movement towards the display could be used to zoom, and movement away from the display could be used to pan~\cite{radle2013effect}.  

Finally, personal devices fitted to the user's arm, like wrist-worn displays or smartwatches~\cite{horak2018david}, can be used both distantly as a remote controller or as an extra input modality while directly touching the LHRD from close range. One example is SleeD, a personal sleeve display showing local visualization lenses or detailed information on the arm, where the finger of that arm is indicating the position or data mark of interest directly on the screen~\cite{von2014sleed}.

\subsubsection{Mid-air gestures}

Free-hand mid-air gestures offer a promising alternative to interact with LHRDs without using additional input devices.
Unlike touch interaction, users can stand at a distance and move freely in front of the display~\cite{vogel2005distant}. 
This approach relies on the postures and motion of the user's bare hands, which may be detected using a depth camera~\cite{yoo_dwell_2015} or through an inertial measurement unit (IMU) and electromyography from an armband device~\cite{haque2015myopoint}. 

Mid-air gestures in front of LHRDs are often similar to touch gestures on smaller devices, but they use more physical space~\cite{wittorf2016eliciting}.
Mid-air gestural interaction is generally slower than direct touch, but performs relatively well to reach larger targets at much farther distance.
Techniques using fingers are faster than those requiring hand or arm movement in 3D space~\cite{nancel2011mid}.
Holding the hand and fingers in a static mid-air pointing posture is more straining and tiring, which might lead to the gorilla-arm effect~\cite{jakobsen2015should}.
``A relaxed arms-down position with both hands interacting at the sides of the body'' may ease the interaction with LHRDs~\cite{liu2015gunslinger}.

Beyond target \emph{selection}, more interaction tasks can be performed like \emph{exploring}, \emph{reconfiguring} the objects in the workspace, and \emph{abstracting/elaborating} ~\cite{matulic2018multiray}.
Malik et al.~\cite{malik2005interacting} argue that direct \emph{selection} and \emph{reconfiguration} tasks are better suited to vision-based hand tracking interfaces due to their low learning curve than systems using complex gesture sets, as proposed by Kjeldsen and Hartman~\cite{kjeldsen2001design}.
Hand and posture mappings can be linked to perform various interaction tasks. For instance, the number of fingers may map different interaction intents: one finger for \emph{select} and \emph{explore}, two for zoom to \emph{abstract or elaborate}, and four to \emph{undo-redo}~\cite{liu2015gunslinger}. 
In the Multiray approach~\cite{matulic2018multiray} each finger projects a ray onto the screen, where patterns of ray intersections formed by hand postures generate 2D geometric forms to perform more elaborate tasks than selection. For example, a lens widget can be triggered by forming a circle that can be moved around or rescaled by bimanual hand gestures~\cite{kister2015bodylenses} to \emph{abstract and elaborate} on various regions of the visualization. 

\subsubsection{Gaze}
As eye gaze follows a user's focus, eye movements can be used to carry out visualization tasks on LHRDs.
Since eye gaze is ``always on'', one must provide a mechanism to avoid unintentional interaction also known as the ``Midas Touch'' problem~\cite{stellmach2013still,lander2015gazeprojector}. Since there are more pixels and more space in an LHRD setting, gaze input can be used to analyze the visual behavior of the user which can provide a deep understanding of gaze patterns ~\cite{chuang2010measuring}. 

Real-time eye gaze may serve various interaction tasks.  The cursor can be associated to eye gaze to \emph{select} an item quickly or to \emph{reconfigure} the interface by moving the items displayed on the LHRD. The user may select an item by looking at it, then drag it quickly across the screen~\cite{stellmach2013still}.
It is also possible to \emph{elaborate} on dense regions of the visualization to reveal details based on eye gaze or to \emph{encode} data differently and highlight recently visited spots~\cite{herholz2008libgaze}. Eye gaze information can be used to automatically \emph{reconfigure} the elements on the interface or to \emph{filter} the visualization.

\subsection{Utilizing the Space in Front of LHRDs}
\label{subsec:spatial-front-LHRD}

Users of LHRDs often alternate between a close position and a distant observer position, thereby using the space to implicitly or explicitly control the visualization. We can distinguish whether the visualization is actively changed due to that movement (Proxemic Interaction) or basically remains unchanged (Physical Navigation). In addition, position and orientation of mobile devices that are tracked in space in front of the display can be used for spatial interaction, as already mentioned in section \ref{subsec:remote-controllers} on remote controllers.

\subsubsection{Physical navigation}
The size of LHRDs allows for physical navigation \cite{ball2007move}, exploiting user movements and viewing direction for navigating an information space. 
Users can interact with the data displayed and reach targets visually by moving their body or turning their heads (the view remains the same), unlike mouse-based pan and zoom interactions in regular desktop environments (the view is changed). 

Physical navigation supports two interaction tasks: \emph{explore} and \emph{abstract/elaborate} the content of the visualization. Moving from one side of the display to the other at a constant distance from the display is an example of exploration amounting to a physical panning interaction. To physically zoom, the user can simply walk closer to or away from the physical display. Physical navigation has been shown to speed up search tasks in a map visualization by more than tenfold~\cite{ball2007realizing}. Search performance in corpus visualizations also benefits from physical navigation~\cite{lischke2015using}. 
Ball et al. found that users prefer physical navigation to virtual navigation~\cite{ball2007move}.

This performance can be linked to greater use of spatial memory when moving in front of the display~\cite{jansen2019effects,radle2013effect}. Such a benefit must be supported by visual \emph{encodings} that promote this movement. Encodings should consider visual aggregation and perceptual scalability when the user moves away from the display since it clearly impacts  performance~\cite{yost2007beyond,endert2011visual}. The hybrid-image visualization~\cite{isenberg2013hybrid} is an example of a successful consideration of these aspects. As described in section~\ref{sec:time-vis}, the idea is to blend into a single static view different representations, each being visible from a specific distance to be assumed via physical navigation.

\subsubsection{Proxemic interaction}
Proxemic interaction exploits users' physical movement as an input~\cite{greenberg2011proxemic}.
Various dimensions of movement can drive the interaction with a visualization: distance, orientation, speed, identity and location~\cite{jakobsen2013information}.
Proxemic interaction can be used to \emph{abstract or elaborate} on the level of detail of a visualization based on user position and orientation. Lehmann et al.~\cite{lehmann2011physical} dynamically expand or collapse nodes of a hierarchical graph visualization based on the user's position in discrete zones in front of the LHRD.
The proximity to the display can also determine the type of visual \emph{encoding} to be displayed~\cite{dostal2014spidereyes}.
As a kind of meta interaction, Kister et al.~\cite{kister2015bodylenses} automatically adjust the mode of interacting with a visualization. They use coarse-grained interaction with free-hand gestures distantly and direct touch interaction close up.

Proximity and movement can also be used to help users to \emph{select} data points in the visualization.  
For example, in the case of multilevel data structures, such as geographic maps, moving back and forth from the display allows the user to access different levels, such as country, city, or district and interact with everything that falls inside the visualization~\cite{peck2009multiscale}.
Body orientation may also be used to indicate areas of interest, allowing users to \emph{connect} data points across these areas and highlight them or \emph{reconfigure} the data and sort them based on a particular variable~\cite{jakobsen2013information}.
Views and legends could be automatically \emph{reconfigured} depending on user movement in front of the LHRD to adjust to the user focus~\cite{jakobsen2013information, Radloff11SmartViews}.

Proxemic interaction raises various design opportunities including controlling a lens with one's body~\cite{kister2015bodylenses}, providing the user with a container and private area~\cite{shoemaker2010whole,shoemaker2010body}, and allowing users to visualize workspace awareness cues by displaying role-specific data~\cite{prouzeau2018awareness}.

Ultimately, for example in ``be the data''~\cite{Chen2018-TLT}, proxemics might be used to teach students high-dimensional data analysis.
Using proxemics, each student embodies a data point in the system. Students physically walk about the room in relation to one another in order to collectively generate interesting insights in data analytics and obtain visual feedback on important data dimensions.

\subsection{Multi-display Interaction}
\label{subsec:MDE-interaction}

Smaller display devices may be used to interact with the data displayed on an LHRD, or to show a user a different level of detail or graphical representation than those shown overall on the LHRD, or to display restricted data to authorized users only. 

\subsubsection{Mobile devices}
Besides their possible use as eyes-free remote controllers, 
smartphones and tablets may be used as an extra display to carry out visualization tasks in LHRD environments~\cite{kister2017grasp}.
For example, Smarties~\cite{chapuis2014smarties} is a mobile application for smartphones and tablets which allows users to control one or more cursors on LHRDs. Cursor positions can be shared between users. All cursors are visible in the mobile application, which supports awareness and collaboration. 
Similarly, SketchSliders~\cite{tsandilas2015sketchsliders} allows users to freely sketch custom sliders on mobile devices and use them to \emph{filter} the data based on certain dimensions or \emph{reconfigure} the plots on LHRDs.
To free users from carrying additional mobiles, arm-mounted devices such as SleeD~\cite{von2014sleed,von2015using} or lightweight wearable smartwatches~\cite{horak2018david} can be used instead.
For example, one may \emph{select} a part of a map on an arm-mounted mobile device and display it on the LHRD, or may \emph{filter} the data displayed on the LHRD using the controls displayed on a sleeve display~\cite{von2014sleed}.
Anyway, attention switching problems may arise as the user's gaze switches between the LHRD and the handheld or wearable device.

\subsubsection{Head-mounted displays}
The combination of see-through Head-Mounted Displays (HMDs) with LHRDs is very appealing for co-located,
collaborative data exploration. 
Prior work combined LHRD with see-through HMDs to interact with graphs~\cite{sun2019collaborative}, volumetric data~\cite{nagao2016enabling,Alsaiari2019-SMC}, and multivariate data in MCV~\cite{Reipschlaeger2020-TVCG}. 
The LHRD is used to share information among a group of users while the HMD provides users with augmented reality views as individual private spaces, for example, to test hypotheses without polluting the shared space and only share interesting findings afterwards.
The \emph{encoding} of information on LHRDs, the \emph{reconfiguration} of objects layout, and the \emph{filtering} of information may change when information is shared. HMD users may have role-based privileges, to access and \emph{elaborate} details concerning the displayed information on LHRD, not granted to the rest of the group~\cite{sun2019collaborative}. \emph{Exploring} and interacting with additional private views can be based on user interest, spatial position, and role. 
HMDs can leverage 3D perception and give access to mixed-reality layers and different visual \emph{encodings}. 
Also, when a user \emph{selects} an item in a visualization displayed on the LHRD, the related information is highlighted in the HMD~\cite{Reipschlaeger2020-TVCG}.

\subsubsection{Multi-device ecologies}
Multi-device ecologies consist of many devices of various sizes and purposes, including LHRDs.
They aim to create novel workplace environments supporting user interaction across multiple displays; for an overview see the cross-device taxonomy by Brudy et al.~\cite{Brudy19Cross-Device}.
Cross-device interaction occurs in smart rooms when multiple coworkers can seamlessly use many displays of varying sizes to achieve a common goal. Each kind of display plays a different role. LHRDs can be used to provide an overview and share information among users, such as in traffic control centers. 
Tabletops may serve as semi-public interaction areas, whereas personal displays, e.g., workstations, tablets, smartphones, smartwatches, and smart glasses, provide role-specific interactions~\cite{chokshi2014eplan,prouzeau2018awareness,radloff2015supporting, Alsaiari2019-SMC}.

\subsection{Multi-modal Interaction}
\label{subsec:multi-modal-interaction}

All previous interaction approaches have their pros and cons, and any interaction design usually needs to consider trade-offs (e.g., between precision and naturalness). That said, multi-modal interaction aims to provide the user with the best of several approaches and strives to compensate for disadvantages.

One possibility is to let users choose freely and naturally between different modalities according to their current position anywhere in the space. For example, the user can use direct touch on the LHRD in near-mode and mid-air gestures or handheld devices in far-mode~\cite{bragdon2011code,langner2018multiple}. Also body movements, freehand gestures, and touch and pen interaction have been combined for lens manipulation on an LHRD~\cite{kister2015bodylenses}.

Different modalities can also be combined seamlessly. For instance, the user may \emph{reconfigure} an object by physically \emph{selecting} and touching it on the display, and then continue dragging the object after transitioning to mid-air interaction without having to explicitly switch between the two modalities~\cite{rateau2016talaria,arslan2019pad}. 
Other modalities that have been combined for data visualization -- though not always for LHRD -- are pen and touch interaction~\cite{Kim:2019:Inking, Romat:2019:Th-Inking, Frisch:2010} or speech and touch~\cite{Srinivasan:2020}.
Yet, we still lack synergistic multi-modal interaction approaches for data visualization, especially for LHRD.

The choice of a modality often a trade-off between speed and accuracy. 
Dual-mode techniques provide a coarse mode allowing to quickly \emph{select} a region of interest, and a precise mode to \emph{select}  a target in this region.
For example, gaze or head movement were used for the coarse mode while touch was used to make precise adjustments~\cite{stellmach2013still,nancel2013high}.
Mid-air gestures, while being error-prone and slow when selecting small targets, are nonetheless suited for coarse interaction, whereas speech~\cite{tse2007speech} and on-body touch~\cite{wagner2013body} support precise selections.

LHRDs can also exploit both explicit and implicit interactions. 
An interaction is implicit when the user's primary goal is not to interact with the LHRD when she moves, turns or steps closer/farther to the LHRD. 
An interaction is explicit when the user's primary goal is to interact with the information presented on the LHRD, e.g., hand swiping as a form of mid-air gesture.
For instance, a combination of implicit and explicit interaction could be used for various types of interactive lenses: zoom and filter, scale, merge, split, etc.~\cite{badam2016supporting}.

\subsection{Multi-user Interaction}
\label{subsec:multi-user-interaction}
LHRDs are useful for co-located collaborative work. 
As the workspace can be shared between multiple users, new opportunities for multi-user cross-device interaction arise. 
Sadly, user interaction may impact the entire LHRD and disturb coworkers or lead to interaction conflicts when coworkers compete for the same items.
One solution is to build socially translucent systems satisfying the three properties of Erickson and Kellogg's framework: visibility, awareness, and accountability~\cite{erickson2000social}, such as displaying awareness bars to show the focus of other users on the edge of an LHRD or magic-lenses showing role-specific data~\cite{prouzeau2018awareness,kister2015bodylenses}.

Avoiding interaction conflicts usually requires user identification. If a system can distinguish the touch, pen, or other gestural/body input of individual users, it is far easier to support conflict-free loosely coupled collaboration or just parallel work. User identification on LHRDs is still a research topic with few solutions, e.g.,~\cite{von2016youtouch}.

Shared interaction techniques where each user carries out a part of a common command or task~\cite{liu2016shared,liu2017coreach} have also been studied. The first user may \emph{select} an item to be dropped elsewhere in the LHRD by a second user. This boosts collaboration even when users are far apart, reduces physical navigation, improves operation efficiency, and provides a more enjoyable experience. 
Looking at how pairs of users collaborate, Prouzeau et al.~\cite{prouzeau2016evaluating} designed two techniques to tackle graph exploration tasks in multi-user touch scenarios on LHRD. Users split the LHRD spatially even when the tasks were not clearly divided. Yet, Prouzeau et al. noted a trade-off between awareness of other users' work and visual disruption when a user makes changes that may affect the partner's work.

Also, Knudsen et al.~\cite{Knudsen2019-VAHC} established a set of interaction mechanisms for multiple views, like view cloning or view creation from existing views, to promote rapid and flexible collaborative data exploration on big screens for healthcare data analysts' work using touch.
Langner et al.~\cite{langner2018multiple} investigated collaboration of pairs of users with MCV of multivariate data at LHRDs and found that users consider movement positively, often move and vary their distance to the display, and stand and walk close to each other often.

In brief, this section covered various interaction possibilities based on the seven interaction tasks of Yi et al. Much work has used novel interaction modes, e.g., cross-device interaction, proxemics, and physical navigation to exploit the space around LHRDs. Various collaborative interaction techniques were also explored. It is yet hard to select the most suitable interaction technique, due to the lack of accepted design guidelines, and the diversity of goals (e.g., boosting collaboration, increasing reachability, reducing interaction time, avoiding interaction conflicts). Also, evaluating interactive visualization on LHRD requires specific strategies, discussed next.

\section{Evaluation Strategies }
\label{sec:studies}

After surveying visualization and interaction on LHRDs, we now review how they have been evaluated, including the goals, questions, types of results, and methodological approaches. We rely on the seven scenarios of visualization evaluation of Lam et~al.~\cite{lam2011empirical}:

\begin{enumerate}[noitemsep]
    \item Understanding Environments and Work Practices (UWP)
    \item Visual Data Analysis and Reasoning (VDAR)
    \item Communication through Visualization (CTV)
    \item Collaborative Data Analysis (CDA)
    \item User Performance (UP)
    \item User Experience (UE)
    \item Algorithm Performance (AP)
\end{enumerate}

This categorization provides a high-level overview of evaluation goals and a useful perspective on reasons and assumptions in evaluation.
Below, we recall the definition of each scenario, describe its presence in our corpus, and discuss, in the context of LHRDs, \emph{what} is tested and \emph{how}, with examples. In \autoref{app:tables}, \autoref{tab:evaluation_questions_examples} provides an extended list of research questions for each evaluation scenario.

\subsection{Understanding Environments and Work Practices}

Prior to designing solutions for LHRDs, user studies must be run to better understand the work practices of the target users.
The studies may assess whether such work practices can be linked to specific benefits ascribed to the use of an LHRD.
For example, it is useful to tell if an LHRD has an added value for certain user tasks, especially with expensive resources, e.g., office space or qualified staff. 

In the early phases of LHRD-based system definition, user experience methods are used to map out the key characteristics of the system: what it is, who it is for, and the usage context. Study methods include interviews, field and laboratory observations. Outputs are often narrative accounts of workflows, work practices, thought processes of users and the underlying structure of their activities when using their current tools or displaying their data on an LHRD.

For example, the requirement gathering phase of the BactoGeNIE system, a comparative genome visualization for LHRD~\cite{aurisano2015bactogenie}, took two years during which the authors ran an ethnographic observation, interviews, and focus groups with eight genomics researchers. Similarly, Wigdor et al. ran a one-year long ethnographic study (intensive interviews, observation group meeting study) to understand the workflow of a group of researchers for a visual collaboration system on an LHRD~\cite{wigdor2009wespace}. To spark new ideas for using LHRDs, Liu et al. brought experts from different domains in front of an LHRD to look at relevant data and arrange them together~\cite{liu2017coreach}. Rajabiyazdi et al. ran contextual semi-structured interviews in front of an LHRD by showing to researchers their own data to understand the potential and thorough use of the technology~\cite{rajabiyazdi2015understanding}. The US National Fusion Collaboratory Project also ran an observational study to identify the best ways of using an LHRD to support collaboration in control rooms~\cite{abla2005shared}.

\subsection{Visual Data Analysis and Reasoning}

This scenario measures users' ability to understand complex data sets and to explore the data from multiple perspectives using visual analytics interfaces deployed on LHRD compared to, e.g., a regular display. 
User studies may assess the contribution of LHRD-based visualization to the analytic process. 
They often compare the quantity and quality of user insights in LHRD and desktop environments.
A known challenge here is that user interfaces (UI) built for desktops have quite different design assumptions, which makes their use as-is on an LHRD unlikely to exploit its benefits. The spatial layout and interaction modalities of the UI may require a major overhaul. 

Field studies, often in the form of case studies, constitute the most common assessment method in this category. Outputs include both quantitative metrics like the number of insights found during the analysis~\cite{reda2015effects} and subjective feedback like comments on the quality of the data analysis experience~\cite{reda2012scalable, rajabiyazdi2015understanding}. 

For example, to study the role of LHRDs in the sense-making process, users were asked to solve an analytic problem from a VAST challenge~\cite{andrews2010space}. Researchers may rely on observations to examine the actual mechanics of sense-making. The diary method was used too to electronically record various aspects of the data analysis task like thought process and results~\cite{ruddle2013leveraging}. Reda et~al.  focused on the analysis of the video and audio data recorded during data exploration. They scored user insights and hypotheses~\cite{reda2015effects}. Another method to gather user thoughts on a task is to conduct a semi-structured interview shortly after the task is completed~\cite{isenberg2009coconuttrix}.

\subsection{Communication through Visualization}

A third scenario evaluates the communicative value of a visual representation regarding goals like teaching, learning, idea presentation, or casual use.
Such studies gauge the gains in terms of communication and users' accuracy at interpreting the information supplied, or their ability to find a data item. This may be done in the context of a single interface or for the purpose of comparing several variations of the same data representation, e.g., different representations on an LHRD, or the same representation on LHRD and other devices.  

The applied methods might be quantitative such as controlled experiments, qualitative via interviews and observation, or a mix of both. Study outputs include quantitative metrics like accuracy, learning rate, and retention, and qualitative feedback like comments.

For example, Anslow et al. asked users to answer questions about software visualizations shown on an LHRD and measured their accuracy. They ran an exit survey to collect feedback~\cite{anslow2010user}. Also, Yost et al. measured response time and accuracy for a set of questions in three visualization conditions~\cite{yost2006perceptual}. Horak et al. also ran a controlled experiment comparing an LHRD $+$ smartwatch condition, to an LHRD only condition for visual analysis tasks. The users had to answer questions in limited time, to find out whether adding a smartwatch improved communication~\cite{horak2018david}.
Effectiveness was measured by the number of correct answers in a limited time, whereas response time measured efficiency.

\subsection{Collaborative Data Analysis}

As collaboration is a key concern for visualization on LHRDs, it is crucial to understand to what extent a data visualization tool supports data analysis in groups.
To this end, researchers study user related aspects as they perform collaborative tasks in front of an LHRD, e.g., information management, territories, collaborative position patterns (i.e., users' physical arrangements and standing positions), user behaviors, and how collaboration interaction impacts the efficiency of sense-making using a mix of devices and modalities.

Collaborative data analysis on LHRD is often evaluated using log analysis or observation during visual analysis tasks. LHRD rooms are often equipped with a motion tracking system, marker-less gait capture cameras for body joints, or position tracking of personal devices. The audio and video of the session, and the screen content of the LHRD with various interaction events, may be recorded. 
Outputs in this category include quantitative indicators like frequency and distribution of interactions, position patterns, proximity, location, movement, and physical demand~\cite{bradel2013large,jakobsen2012proximity,altarawneh2015collaborative}, and qualitative indicators like teamwork and verbal communication, visual attention, and interaction conflicts~\cite{hawkey2005proximity,jakobsen2014up,langner2018multiple}.

For instance, Langner et al. analyzed qualitative and quantitative parameters of collaborative tasks, including observed teamwork, verbal communication and distance between coworkers, and their impact on collaboration styles and effectiveness~\cite{langner2018multiple}. Beside user positions, Jakobsen et al. studied the use of screen space among users~\cite{jakobsen2014up}. 
Similar work analyzed user positions to make findings about position patterns~\cite{altarawneh2015collaborative}, awareness of each other's activities~\cite{jakobsen2016negotiating}, shared interaction~\cite{liu2016shared}, and behavior~\cite{bradel2013large}. 

\subsection{User Performance}

In the fifth evaluation scenario, studies measure objectively how various factors affect user performance,
focusing on a single interactive or visual technique, not an entire visualization solution. 

Controlled experiments are run when a precise hypothesis can be directly tested through empirical studies and reported on with statistical significance tests. Log data are used to capture the values of dependent variables, often time and accuracy. Outputs include quantitative metrics, mostly task completion time and accuracy.

For instance, Ball et al. study the impact of display size on completion time for navigation and search tasks~\cite{ball2007move}. Prouzeau et al. compare two techniques for traffic visualization on LHRDs: DragMagic and MultiViews and assess their impact on task completion time~\cite{prouzeau2016towards}.
Prouzeau et al. also study a novel propagation-based selection technique for graph exploration on LHRDs against a basic technique~\cite{prouzeau2016evaluating}. Several studies compare new techniques designed for LHRD environments to baseline techniques commonly used in desktop environments, for example, drag-and-drop vs. drag-and-pop~\cite{baudisch2003drag}, a window manipulation layer interface vs. a desktop style interface~\cite{rooney2012improving}, virtual navigation vs. physical navigation~\cite{radle2013effect}, and touch vs. mid-air interaction~\cite{jakobsen2015should}. 

\subsection{User Experience}

Beyond measuring time and error, an evaluation can also elicit subjective feedback and opinions about a visualization.
User experience studies can yield insights into the user's thoughts, feelings, needs, attitudes, and motivations when using an early design sketch, a functional prototype, or a finalized product in the context of LHRDs.

The choice of research methods depends on the development stage of the system and the purpose of the study. Formative usability evaluations and expert reviews are suitable for gathering feedback to improve the design. Outputs include subjective quantitative metrics, such as perceived effectiveness, perceived efficiency, perceived correctness, satisfaction, trust, and features liked/disliked. Moreover, qualitative feedback can be gathered through open-ended questions.

Formative studies have measured the acceptance and usefulness of using a sleeve display to interact with an LHRD~\cite{von2014sleed}, and the usefulness of implicit and explicit interaction styles~\cite{badam2016supporting}. 
Prior to running a summative study, a preliminary expert review might be conducted to discover issues in an iterative development process~\cite{Riehmann2020-CGF}.
Several studies hold exit interviews or questionnaires after controlled experiments to better understand the motivations, thoughts, and attitudes of participants~\cite{ion2013canyon,Reibert2020-PACM}. 

\subsection{Algorithm Performance}

Lastly, an evaluation study can aim to measure the performance of a visualization algorithm.
For LHRDs, such studies might focus on rendering many data points or on challenges arising from usage contexts specific to LHRDs. For example, while a desktop display may show a few views, large displays can show many more~\cite{knudsen2016view,langner2018multiple}, which may require to efficiently compute view layouts. 

LHRDs are not turnkey solutions. They are high performance real-time parallel visualization systems, which requires significant software developments to orchestrate heterogeneous hardware and software and exploit their full potential (e.g., pixel streaming, distributed rendering). Evaluation studies typically run benchmark tests to examine the performance of algorithms or the overall system.
Outputs of such studies include quantitative metrics such as rendering time, frame rate, CPU load, or RAM usage.
Theoretical analyses of algorithmic performance are hardly found in the context of LHRDs.

Hung et al.~\cite{chung2013survey} surveyed software frameworks devoted to the development of applications for LHRD and algorithm performance, e.g., rendering engines and middleware for multisurface applications. For example, Tuoris~\cite{martinez2020tuoris}, a framework for visualizing dynamic graphics (e.g., hive-plot, 3D content, SVG maps) was used to benchmark visualization algorithms in various settings.

Overall all seven evaluation scenarios of Lam et al.~\cite{lam2011empirical} have been considered in the context of LHRDs. Yet, we mostly see individual customized studies. Establishing an evaluation framework for interactive visualizations on LHRDs remains an open challenge.

\section{Applications}
\label{sec:applications}

In this section, we review interactive visualization on LHRDs from an application perspective and how the different applications use LHRDs.
This can inspire how this technology may be useful for other application domains.
It also helps to analyze how the type of task guides the choice of visualization and interaction techniques.

\subsection{Large-scale Data Exploration and Analysis}

\begin{figure}[t]
\includegraphics[width=\linewidth]{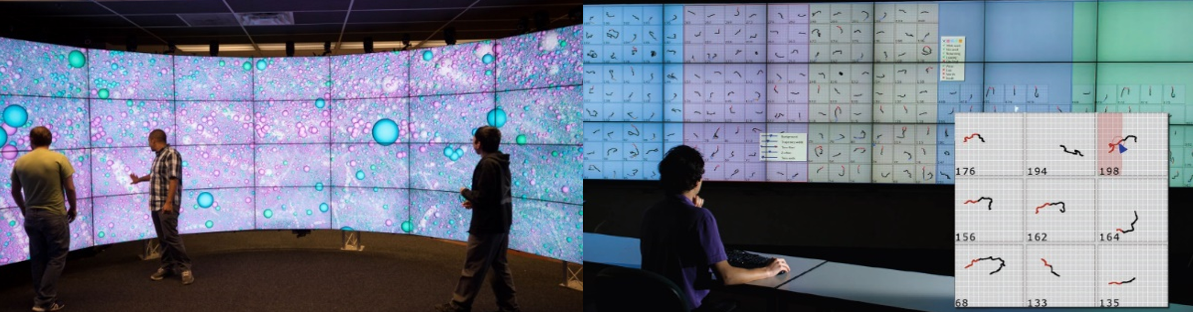}
\caption{LHRD-based visualization of large data. (left) Immersive exploration of a representation of the universe~\cite{marai2019immersive}; (right) visualizing and making sense of ant trajectories~\cite{reda2013visualizing}.}
\label{fig:appLargedataAndSensemaking}
\end{figure}

The primary indication for using an LHRD is to make sense of large data sets.
Analysts can explore interactively a single gigapixel overview visualization a large data set produced by scientific instruments, like astronomy data sets~\cite{pietriga2016exploratory} (\autoref{fig:appLargedataAndSensemaking} left), molecular interactions~\cite{reda2013visualizing} (\autoref{fig:Vis_Physical} left) or genome-sequencing data~\cite{ruddle2013leveraging} (\autoref{fig:Vis_HighRes} left), and 3D/4D geology data~\cite{johnson2006geowall}.

Other examples consist in exploring multiple complementary and feature-rich data sources jointly, such as large multidimensional crime data~\cite{reda2015effects,langner2018multiple}, news data~\cite{andrews2010space}, people and page relationships in social media~\cite{kister2017grasp}, sensor or camera data in a building over time~\cite{badam2016supporting,tsandilas2015sketchsliders}, house sales~\cite{ball2007move,ball2008effects} or insect trajectory data~\cite{reda2013visualizing} (\autoref{fig:appLargedataAndSensemaking} right).
New and better insights were reported in different disciplines when researchers had the opportunity to explore their data on an LHRD~\cite{rajabiyazdi2015understanding}.

\begin{figure}[t]
\includegraphics[width=\linewidth]{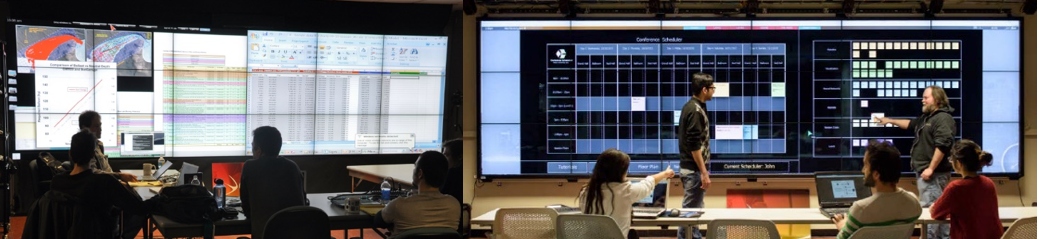}
\caption{LHRDs for workshops and meetings. (left) A multidisciplinary team uses an LHRD to plan the mission of an autonomous vehicle~\cite{marai2019immersive}; (right) an organization committee uses an LHRD to schedule a conference using a multi-user application~\cite{doshi2017stickyschedule}.}
\label{fig:appWorkshopAndMeeting}
\end{figure}

\subsection{Workshops and Meetings}

Workshops and meetings are all about discussion and collaboration. 
LHRDs have been adapted to replace sticky notes and whiteboards usually found in conference organization settings~\cite{liu2014effects}. 
Users liked the visual scalability exposing many aspects of the conference scheduling problem. 
They also found it easier to visualize and manage scheduling issues than with a whiteboard~\cite{doshi2017stickyschedule} (\autoref{fig:appWorkshopAndMeeting} right).
Beyond showing scheduling constraints, the added value of an LHRD is most likely in the support for collaboration.

Many industries rely on collaborative problem-solving.
Experts look at the same data to compare and discuss possible courses of action and related outcomes (\autoref{fig:appWorkshopAndMeeting} left). 
LHRDs increase the perceived efficiency of a workshop, the sense of participation, the motivation and the sense of ownership~\cite{nolte2016collaborative}.
For example, it is helpful to use an LHRD in agile software development to discuss a complex software project containing a collection of digital objects such as classes, packages, and different dependencies. 
The users can select, explore, filter and connect these objects for many tasks such as bug triage or assigning  work items~\cite{anslow2010user,bragdon2011code,mateescu2015awall}. 
Other examples include supporting business process modeling~\cite{nolte2015towards,nolte2016collaborative} or automotive design~\cite{khan2009toward}.  

LHRDs also help to break out of the classic presenter-audience setting where only the presenter can share material with the audience.
For instance, multiple files coming from multiple user devices can be shown and interacted with simultaneously~\cite{radloff2015supporting}.

\subsection{Command and Control}
\label{sec:command_and_control}

\begin{figure}[t]
\includegraphics[width=\linewidth]{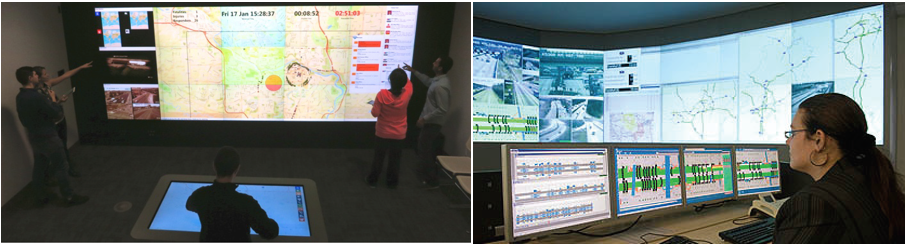}
\caption{LHRDs for command and control. (left) Users collaborate using an LHRD, a tabletop, and personal devices in an emergency response exercise~\cite{chokshi2014eplan}; (right) a dispatcher monitors traffic on an LHRD and sends crews with a desktop PC~\cite{schwarz2012content}.}
\label{fig:appCommandAndControl}
\end{figure}

In command and control rooms too a lot of multi-source information needs to be displayed or cross-analyzed, e.g., traffic or weather forecast data~\cite{prouzeau2018awareness}.
Traffic management in major urban areas relies often on showing overview+detail views distributed on multiple independent displays, e.g., by showing the overview on one display and the details of a focus region on another display. 
In such a setting the analyst's attention is divided between multiple distant displays.
Using an LHRD alleviates divided attention issues by affording enough pixels to display the details of the focus region within its wider context~\cite{schwarz2012content} (\autoref{fig:appCommandAndControl} right).
Concrete applications include the management of traffic lights at about 1500 Parisian crossings, with over 800,000 cars and 2.5 million pedestrians every day~\cite{prouzeau2016towards}, and dispatching law enforcement officers to incident locations visualized in real time on a large map~\cite{ion2013canyon}.

Combined with tablets and tabletops, emergency response rooms fitted with LHRDs allow multidisplinary teams, e.g., from the police, the army, and hazardous materials forces (HAZMAT), to plan and monitor operations collaboratively~\cite{chokshi2014eplan} (\autoref{fig:appCommandAndControl} left). 
The coworkers can visualize multi-source data, e.g., maps and social media data. The LHRD supports information sharing and serves as a shared interaction space between all coworkers~\cite{onorati2015walltweet,abla2005shared}.

\subsection{Health and Medicine}
\label{sec:healthcare}

\begin{figure}[t]
\includegraphics[width=\linewidth]{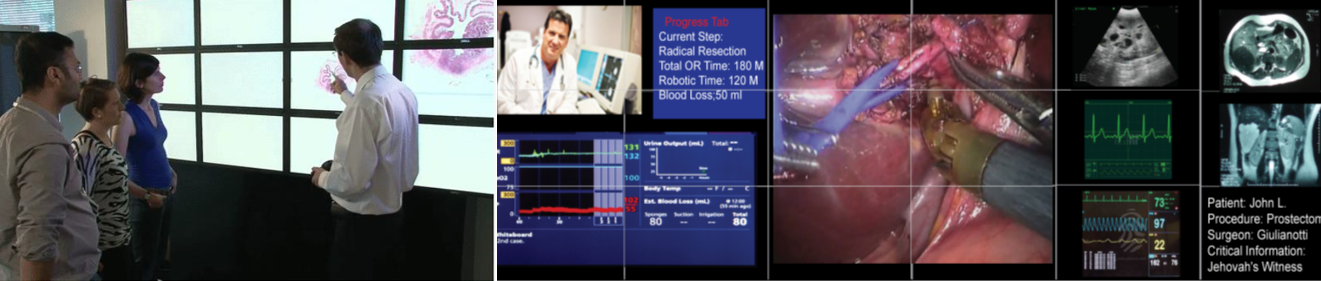}
\caption{LHRDs for healthcare and medicine. An LHRD is used (left) as a virtual microscope to examine tissue sections collaboratively to improve diagnosis~\cite{treanor2009virtual}; (right) during surgery to monitor all relevant patient information and to communicate with remotely located surgeons for advice during surgery~\cite{bhatia2009oproomlhrd}.}
\label{fig:appHealthcare}
\end{figure}

Clinicians need all sorts of information to make decisions about diagnosis and treatment, like patient data, vital signs or medical imagery. 
They need the same to communicate with remotely located surgeons for advice during surgery. 
Prior work has designed LHRD-based solution for such usage scenarios~\cite{bhatia2009oproomlhrd} (\autoref{fig:appHealthcare} right). 
Another avenue is intensive care units to ease the hand-off between teams and ensure the continuity of care~\cite{thomas2017echo} (\autoref{fig:Vis_Temporal} right).

In the area of medical imagery, LHRDs can support two usage scenarios.
The comparative analysis of many high-resolution scan-images helps to compare healthy and unhealthy organ tissues~\cite{gjerlufsen2011shared} (\autoref{fig:Vis_SmallMultiples} right).
The visual analysis of high-resolution gigapixel images helps pathologists to make diagnoses at a much higher magnification level than a typical microscope~\cite{treanor2009virtual,ruddle2016design} (\autoref{fig:appHealthcare} left).

\subsection{Teaching, Learning, and Training}

\begin{figure}[t]
\includegraphics[width=\linewidth]{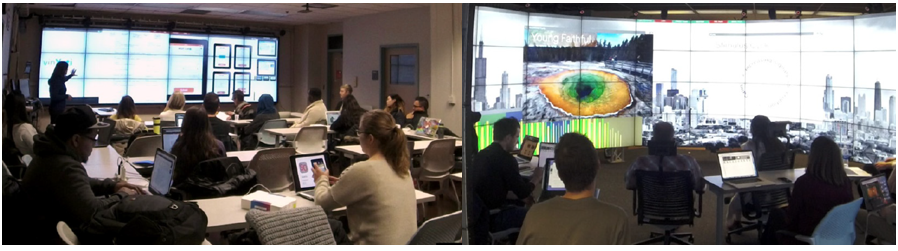}
\caption{Teaching with LHRDs. Two classrooms where a mix of LHRDs and personal devices allow the teacher and students to share and work on documents and take part in discussions~\cite{chattopadhyay2018shared}.}
\label{fig:appEducation}
\end{figure}

Also learning experiences often require the visualization of complex data. With a projector in the classroom, the number of pixels is limited, and the teacher passes slides, while the students are seated in the classroom, passively listening. 
Education is rarely apprehended in an LHRD environment. 
LHRDs could be used to show students patterns in complex data, explain work processes in classrooms, and access some experiences that would otherwise be inaccessible (e.g., access to paintings or items from a museum in another city)~\cite{rajabiyazdi2015understanding}.
In a multi-device environment combining an LHRD with personal devices, students can easily share their documents and talk with well managed turn-taking~\cite{chattopadhyay2018shared} (\autoref{fig:appEducation}). 
Working together in the same space helps students to develop more ideas faster and with higher quality and to be aware of the contributions of others. Students also reported that using an LHRD increased their interest in the content~\cite{clayphan2016wild}.

Like other immersive environments, LHRDs could also be used to train users on simulations of risky situations, e.g., driving and flight simulators, and military or surgery training. Relevant visualizations are similar to those discussed earlier in sections \ref{sec:command_and_control} and \ref{sec:healthcare}.

In sum, we illustrated the benefits of LHRDs in specific application domains or use cases.
Many daily computing tasks such as web browsing, reading papers, programming can become much easier on LHRDs than on desktop environments~\cite{bi2009comparing}.  
We can also use LHRDs to change these daily tasks from an individual or collaborative sequential task to live on the spot group task. An example could be preparing a presentation slide deck or video editing when users may arrange and edit content collectively~\cite{liu2014effects}. 

More work is needed to identify other interesting applications that can only, or much more easily, be implemented with LHRDs than with desktop or mobile devices. We will discuss this opportunity and other research opportunities in the next section.

\section{Research Opportunities}
\label{sec:future}

As we have seen in the previous sections, there is plenty of existing research on interactive visualization on LHRDs. Much of the previous research is mainly technology-driven. Still, developing LHRD visualization solutions and deploying them in relevant application scenarios remains a challenging endeavor. Below, we list promising research directions to advance interactive visualization on LHRDs not only technology-wise, but also conceptually with the goal to strengthen its role as a valuable asset in data analysis scenarios.

\subsection{Display scalability}

Future research may first \emph{develop visualizations that scale with respect to the display size.} Such scalable visualizations would ease the use of data visualizations flexibly across heterogeneous displays.

While display scalability was in the visual analytics research agenda~\cite{Thomas05IlluminatingPath}, it has only recently become an active topic of research. Inspired by responsive web design, the visualization community has begun to study responsive visualization design~\cite{Andrews2017ResponsiveDataVisualisation,hoffswell2020techniques}, so far mostly addressing basic charts and information graphics in interactive news articles. A recent Dagstuhl seminar has looked at responsive design for mobile data visualization, mostly to make visualization work on small-scale devices~\cite{Horak21ResponsiveMobileVis}. Yet, little work has considered the full range of display sizes, from small to regular to big displays -- the work of Radloff et al. on redundant visual mapping being one rare example~\cite{radloff2011}. Making visualizations responsive to the usage context and truly scalable across displays of various sizes seems a promising research direction. Such work should also include space and layout management in terms of visual aspects (how many views, which view layout), interactive control (techniques to reconfigure views), and automation (methods to generate suitable view layouts). Besides scaling existing visualizations, one may also look for new visual representations that are useful for LHRDs. The natural multiscale character of hybrid-image visualizations~\cite{isenberg2013hybrid} could inspire such research.

\subsection{Multi-device visualization}

Another aspect of scalability concerns the ability to use multiple displays in concert~\cite{Brudy19Cross-Device}. The goal is to \emph{create visual analysis environments in which visualization views can span or be distributed on multiple displays.} This allows a multitude of output devices to work in concert with LHRDs, not only to show an even bigger image of the data, but also to ease the analysis of data subsets in more detail on personal devices. This is very useful in collaborative settings to offer personally tailored views, to show detail or lens views without disturbing coworkers, or to use alternative visualization views on interactive mobile devices that are better suited to the user's task.

Current LHRD solutions already display data in multiple views, and related design space analyses describe how these views can be combined~\cite{Javed12CompositeViews,Badam19UbiquitousAnalytics,badam2019-InfoVis,langner2018multiple}. Prior work has also studied how such views can be extracted from an LHRD visualization and shown on a mobile device~\cite{langner2016content,langner2018towards,kister2017grasp,horak2018david}. Smart meeting rooms have also been enhanced with multi-display visualizations~\cite{bragdon2011code,radloff2015supporting, Eichner19MultiDisplay}. Augmented reality (AR) has been used too to provide extra visualization views to individuals on LHRDs~\cite{Reipschlaeger2020-TVCG}. 
Yet, such custom multi-display solutions often do not apply in general. Extracting arbitrary data subsets and creating new views dynamically on different devices remains hard. We need conceptual models and infrastructures allowing many users to extract data and views where and when the task requires it. We need to create partial visualizations on the fly and distribute them across devices~\cite{Horak19Vistribute}, and still track (provenance-wise) and combine them back to a full image of the data. More studies may assess the fitness and effectiveness of specific LHRD$+$devices combinations for visualization tasks and domains.

\subsection{Multi-modal interaction}

Like multi-device visualization, multi-modal interaction is a valuable, but under-studied aspect of interactive visualization on LHRDs. The aim is to \emph{use the best input modalities to ease visual data analysis.} For this, research on interaction for LHRDs has to advance from using one or two input modalities to truly multi-modal interaction.

Section~\ref{subsec:multi-modal-interaction} listed a few bimodal approaches~\cite{bragdon2011code,langner2018multiple,arslan2019pad}. Multi-modal interaction uses more modalities, e.g., touch, pen, tangible, gaze, speech, and proxemic interaction. It is hard to integrate several modalities into a coherent interface supporting data analysis tasks in various ways. The user can choose (or the system can recommend) a way to interact based on the type of task, user preferences, or user position relative to the LHRD. Users can ideally transition between modalities seamlessly, which requires a software architecture that supports smooth modality handover. This would allow users to create complex analytical queries as a mix of, e.g., touch gestures, spoken commands, and body movements in front of an LHRD. Current WIMP interfaces do not often support this kind of interaction, but recent advances on vertical touch surfaces~\cite{Srinivasan21MultiModal} already hint at promising ideas worth extending to LHRDs.

\subsection{Multi-user interaction}

Multi-user interaction should also get a fair share of future work on interactive visual data analysis on LHRDs. The goal is to \emph{enable multiple users to engage in collaborative data exploration and sense-making activities.} Much of the reviewed literature is about showing data on LHRDs, which naturally allows teams to browse the data together. But just browsing is not enough. For a comprehensive analysis, the users also need to interact collaboratively and discuss their individual findings and partial insights to form a coherent big picture of the data. This is where more research is needed.

In Section~\ref{subsec:multi-user-interaction}, we illustrated multi-user interaction on LHRDs~\cite{kister2015bodylenses,liu2016shared,liu2017coreach}. Yet, the step from single-user to multi-user interaction is big~\cite{Mateescu21Collaboration}. Multi-user input faces technical hurdles. Most UIs assume a single interaction focus, i.e., pointer, cursor. Multi-user interaction requires multiple foci, which is unsupported in current software libraries. It also requires to attribute interaction input to the right user~\cite{von2016youtouch}. Also, input from various distances and positions around the LHRD needs to be supported (see above), and individual views provided, e.g., with individual lenses~\cite{badam2016supporting,kister2015bodylenses} or AR overlays~\cite{Reipschlaeger2020-TVCG,James20Personal+Context}.
There are also semantic and social challenges. For example, how should we handle concurrent interactions during parallel work that would lead to conflicting states of the visualization, or how can truly cooperative group interaction be moderated via a suitable collaborative interface for increased analysis efficiency~\cite{prouzeau2016evaluating}? How can the system support various user roles, detect or prevent social tensions during a multi-user data analysis session?

\subsection{Models, taxonomies, and guidelines}

As future research offers new solutions for interactive visualization on LHRDs, we also need to conceptualize LHRD-specific models, taxonomies, and guidelines. Such research will \emph{strengthen the theory} behind interactive visualization on LHRDs. While general theoretical work on visualization and visual data analysis are now commonplace (e.g., task taxonomies, design spaces, conceptual models), specific theoretical work on LHRDs is still scarce.

While the visualization literature offers a solid understanding of tasks through both empirical and conceptual work, we know relatively little about visualization tasks in the context of LHRDs and collaboration. For example, the oft-cited typology by Brehmer and Munzner~\cite{Brehmer2013-TVCG} does not ``explicitly address collaborative use of visualization''. Thus, we lack good understanding of, e.g., task planning, coordination and interpretation among coworkers, and how such activities relate to prior task descriptions. More theoretical work could investigate task taxonomies specifically for LHRDs, focusing on gaps in current taxonomies, e.g., access to out-of-reach parts of the visualization or collaborative work. In general, we need to better understand how collaborative LHRD solutions can be designed and used, which calls for new (or adapted) design, implementation, and analysis process models.
An interesting question regarding the design is how to bring interactive visualization to LHRDs. In which cases is it fine to adapt existing approaches and when must we design entirely new visualizations? Similar questions are currently also raised regarding visualization on mobile devices~\cite{Lee:2021:MDV2}, which could inspire similar research on LHRDs. Future research on theoretical foundations must also consider the multi-device, -modal, and -user aspects discussed earlier. Such research work should lead to guidelines or rules as practical advice for newly developing or successfully applying interactive visualization to LHRDs.

\subsection{Toolkit and authoring support}

A critical concern when it comes to developing interactive visualization solutions for LHRDs is to master a quite complex technical environment. There are only very few standard libraries or tools that would help developers~\cite{renambot2016sage2}. Future work should therefore \emph{reduce the technical burdens and make implementing LHRD visualizations easier} by providing toolkits and authoring support.

At the device-level, future research could explore architectures and infrastructures for systems that work with multiple users, multiple displays and multiple interaction modalities driven by multiple computers. At the software-level, we need support for integrating more, and more scalable (small to large) data visual visualizations, more diverse ways of interacting with them, and more users operating the system. New toolkits and libraries should abstract away the technical details and diversity across device vendors and operating systems to ease the deployment of visualizations on LHRDs. Authoring tools should also offer best-practice templates and recommend suitable design alternatives. They need to deviate from regular visualization authoring tools by also considering the multi-(device, modal, user) aspects inherent to LHRD and help to adapt to them.

\subsection{Evaluation studies and applications}

Better support for developing interactive visualizations for LHRDs can be a catalyst for more research on design, user, and evaluation studies and more widespread utilization in diverse applications. Such research would help us \emph{develop a better understanding of the advantages and limitations} of interactive visualization on LHRDs.

While LHRDs have been applied in various domains, prior work is mainly technology or research-driven. But is there really a value in LHRD visualization, can it really solve domain problems, and, if so, in what domains is it most beneficial? What is needed are design studies on the utility of LHRDs for a broader spectrum of application domains and ideally also comparative evaluation studies to gain more insight in terms of different display technologies. For example, biological data visualization can benefit from the larger display space of LHRDs~\cite{ruddle2016design}. But virtual reality (VR) also offers a larger (though virtual) display space for biological data visualization~\cite{Ripken21Immersive}. So far, however, biologist will certainly find it difficult to tell which technology (LHRD or VR) to use and for which tasks they are particularly useful. This is where comparative studies could provide some help. Yet, such studies will be challenging to conduct due to the multitude of aspects to be considered, including technological issues, human perception and cognition, collaborative data analysis, and user immersion. As mentioned before, better toolkit and authoring support would at least lower the technical hurdles for such studies. Dedicated evaluation frameworks for LHRDs would make it easier to plan and conduct the studies.

\subsection{Societal impact}

This survey showed that LHRD research has so far been mainly technology-driven. Much of our discussion on future work carried on in a technological direction, also including theoretical questions in the area. We may yet call for more research that transcends technology and theory. Such research would aim \emph{to embed interactive visualization on LHRDs properly in a responsible human society.}

This could first mean to democratize the use of LHRDs. Currently, they are mainly used in research labs or high-profile institutions and companies --- the work of Walny et al. on a tool to ease data engagement events in public spaces being one rare example~\cite{walny2020pixelclipper}. What would it mean to make LHRDs more ubiquitous and support their open access in public spaces like libraries, cultural centers, shopping malls, and museums? People could interact with weather or election data, map visualizations, visualizations in news or educational contexts -- moving from in-depth visual data analysis to casual and lightweight information graphics for all. Thus, many research challenges would arise pertaining to public vs. private use, privacy issues, spontaneous and easy interaction, BYOD support, social interaction around these displays etc. This relates to extensive research in ubiquitous computing, pervasive displays, and digital signage. With wall-sized display technologies becoming cheaper and more popular (like with today`s larger TV sets), LHRDs could also enter private homes. Directly related to a more ubiquitous, personal use of LHRDs is the growing diversity of tasks such displays are used for. Here, the larger display space could also be used to offer extended contextual information to private users looking at their domestic energy consumption, optimizing travels for reducing their carbon footprint, or making better-informed purchase decisions. LHRDs could play a role in going beyond the currently dominant mobile device usage and becoming a valuable addition in future device ecologies. We still need to figure out how to integrate work with a large display with other analytics contexts, or what LHRD sizes are suited for certain room sizes.
Of course, issues like energy consumption and sustainability directly apply to LHRDs themselves. Currently, maintenance of LHRDs is costly, and their lifespans are relatively short with regard to initial investments. The development of future large display technologies might mitigate these problems.

In sum, while there has already been much research on LHRDs, there are still many open questions. Here we suggested several research avenues related to technological, theoretical, and societal aspects. We hope that the listed research topics will inspire future work advancing interactive visualization on LHRDs. We also considered the practical side of running literature reviews, where LHRDs can be a valuable aid in collaborative literature reviews.

\section{Conclusion}

This work reviewed the literature on interactive visualization on large high-resolution displays. Driven by the questions ``Do we truly build interactive visualizations \emph{for} LHRDs?'', we looked in detail at previous research on the visualization of data on LHRDs and on the interaction with such visualizations.
We found that aside from few works, there appears to be a lack of visualization techniques adapted or built specifically for the needs of LHRD situations (more pixels, more space, more users, and more devices).

We further reviewed LHRD evaluation methodologies and application scenarios. In extracting relevant study questions and research methods, we acknowledge the necessity to strengthen the theory behind assessing interactive visualization on LHRDs. We identified application domains and corresponding examples to illustrate how LHRD technology may be used to solve real-world problems. This has the potential to open the door to a greater range of application domains, which would also help future studies to examine the added value of LHRDs through comparative evaluations.

In this review, we discovered several aspects that are not adequately addressed by the existing literature and outlined corresponding opportunities for future research. We are convinced that LHRDs, when paired with cross-device interaction, will play an essential role in many future visualization scenarios, particularly those requiring collaborative data analysis, sense-making, and problem solving. Working on the identified research challenges will foster the adoption of visualization solutions deployed on LHRDs for relevant application scenarios. Our survey can serve as a good starting point for the advancement of this technology.

\bibliographystyle{eg-alpha-doi} 
\bibliography{references}       


\clearpage
\newpage
\renewcommand{\thetable}{\Alph{section}\arabic{table}}
\setcounter{table}{0}
\appendix
\section{Search Methodology}
\label{app:method}

\subsection{Identification and Information Sources}
\begin{figure}[htb]
\includegraphics[scale=0.54]{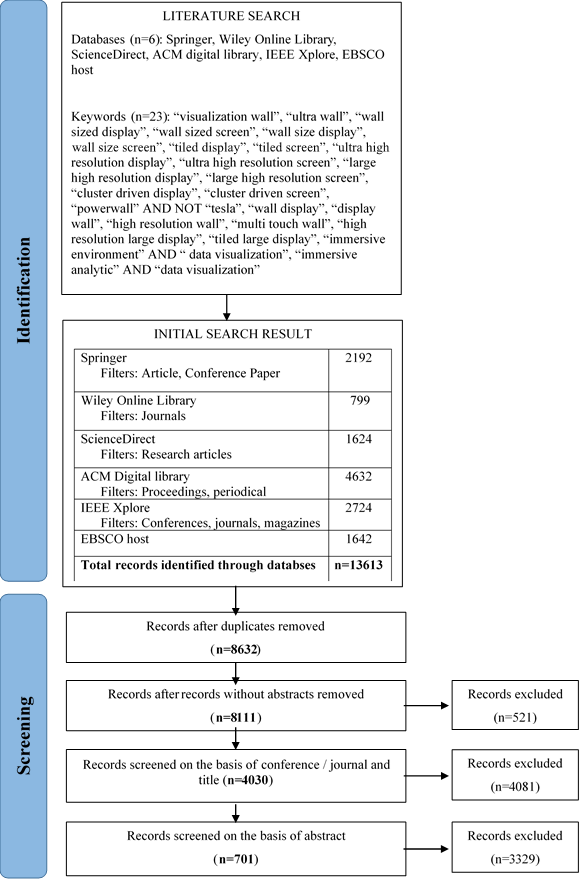}
\caption{Flowchart of our systematic search adapted from PRISMA~\cite{moher2009preferred}.}
\label{fig:prismaMethodo}
\end{figure}

To find relevant papers for our survey, we conducted a systematic search in six major scientific databases in computer science (Springer, Wiley Online Library, ScienceDirect, ACM digital library, IEEE Xplore, EBSCO host between November $23^{rd}$ and December $01^{st}$, 2019) using 23 search queries capturing the different terms used to describe a wall-sized display (e.g. “visualization wall”, “tiled display”, etc.). The methodology steps are summarized in an adapted PRISMA~\cite{moher2009preferred} study flowchart (Figure \ref{fig:prismaMethodo}).

We used general search queries such as (“immersive analytics” AND “data visualization”) and specific keywords like “wall-sized display” to ensure a broad bibliographic search. 
The queries were executed on the full text with filters on conferences and journals while ignoring any other types of publication like book chapters (except for EBSCO-host). We have also entered these queries in singular and plural forms when the search engine did not support lemmatization natively (Wiley Online library, IEEE Xplore and EBSCO-host).

\begin{figure*}[h]
\includegraphics[width=\textwidth]{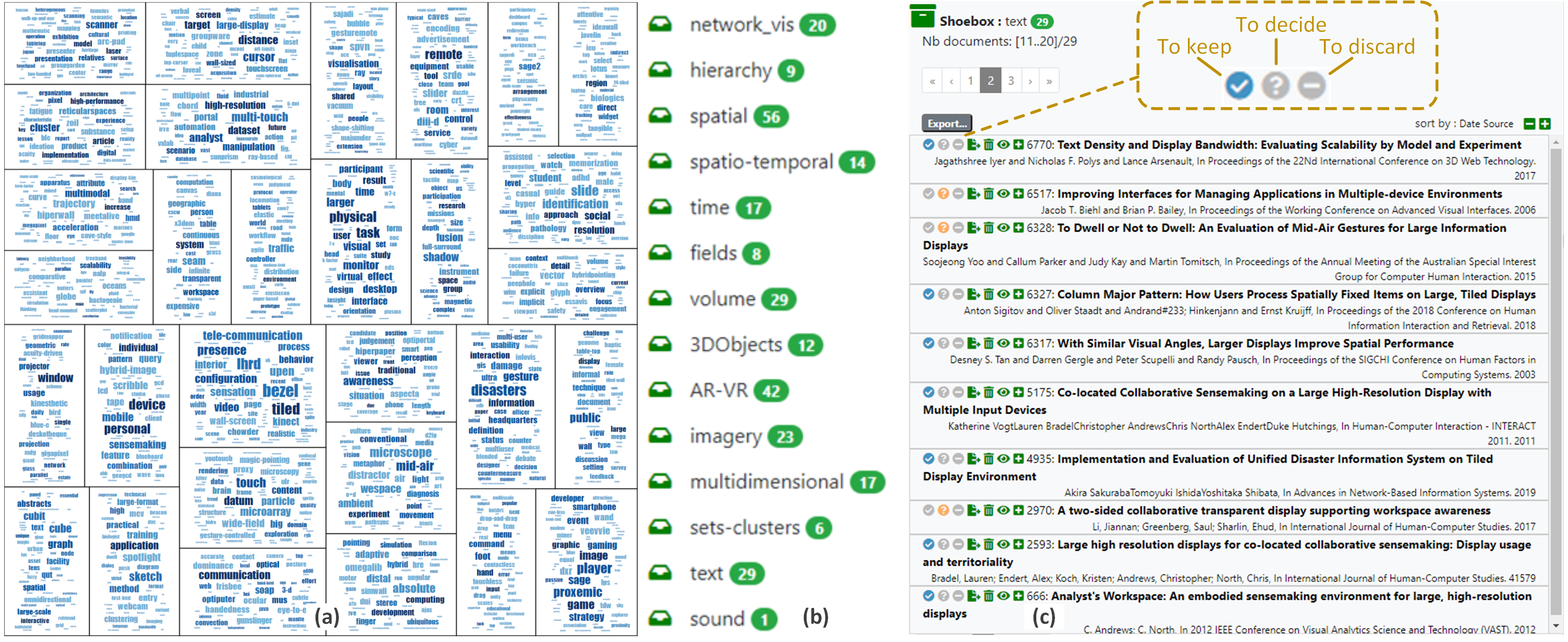}
\caption{Papyrus software~\cite{medoc2016exploratory} showing (a) a map of topics extracted from the scientific corpus of 701 papers, (b) $14$ shoeboxes created with the related number of papers, (c) a subset of papers assigned to the shoebox related to textual data with the annotation mechanism.}
\label{fig:papyrusShoeboxes}
\end{figure*}

We exported all query results to a spreadsheet with the metadata for each record: title, year, DOI, conference or journal name, abstract, authors, keywords, search query and database. We used the "export" functionality in each database when available. We also developed a Web scraper to extract paper abstracts from the ACM digital library and Springer library.

\subsection{Screening and Selection Criteria}

\paragraph{Removing duplicates and papers without abstracts}

The initial spreadsheet included 13,613 records. Some data curation was carried out to fix character encoding issues. After cleaning and merging duplicate entries, the number of records was reduced to 8,632. Then, we excluded papers having no abstracts considering they lacked scientific contribution (e.g. calls for participation, proceedings front matter, and keynotes). 521 records were excluded consequently and a total of 8,111 records were retained.

\paragraph{Screening based on the relevance}

Using a visual text mining tool, Papyrus~\cite{medoc2016exploratory}, to visually inspect this corpus (see Figure~\ref{fig:papyrusShoeboxes}-a), we identified topics that were irrelevant for our literature review e.g. biological conferences where papers discuss the properties of the “cell wall”, or the thermal properties of brick walls in construction journals.
This justified a second filtering step whereby irrelevant journals and conferences, and their papers, were trimmed out of the corpus.

To this end, another round of data curation was required to harmonize and deduplicate conference/journal names. We first deleted year/volume information using regular expressions (e.g. 10th international conference, 2016 international conference, TEI’16, etc.). We also calculated the normalized Levenshtein distance for all pairs of names keeping pairs with similarity greater than 0.8 in order to cope with variations in word order, abbreviations, splitted words, (e.g. “Australian Conference on Computer Human Interaction” and “Australian Computer Human Interaction Conference”, “Universal Access in the Information Society” and “Univers. Access Inf. Soc.”). We manually screened the resulting similarity pairs to remove false positives. Hence, we were left with 2,665 distinct conference/journal names containing the 8,111 records.

We then sorted these names based on the number of records they had, from the most popular (373 papers in the CHI conference) to the less popular (only 1 paper in 1628 conferences/journals). For venues having more than three papers, if at a first glance we found one relevant record title, all records for this venue were retained, otherwise they were excluded. For venues having three or fewer records, if their names sounded relevant to the VIS community and to this survey, we kept their records. In the end, we eliminated 4,081 records and kept 4,030 published in 534 distinct journals/conferences.

\paragraph{Screening based on abstract}

To further reduce the corpus to consider in the survey, we loaded the resulting 4,030 records in Papyrus. Latent topics were automatically extracted and visualized. Drilling down into a given topic, one could also access the list of papers (see Figure~\ref{fig:papyrusShoeboxes}-c) included in it and annotate every paper according to whether it should be kept, discarded or a decision could not be made based on the title and abstract only.
One of the co-authors has screened the 4030 abstracts to annotate them according to the following criteria. Excluded were papers related to hardware, OS/Network protocols, distributed rendering and distributed user interfaces on LHRDs, other devices and technologies (e.g. mobile devices, wearable tangible displays, tabletops) when they are not combined with LHRDs. We kept papers dealing with visualization techniques on LHRDs. As a result, 3,329 records had to be discarded and 701 records were annotated as relevant or requiring to read their full text.

\paragraph{Full text assessing and tagging system}

We used the 701 papers as an exhaustive scientific corpus for our survey. The individual co-authors could use Papyrus software to further filter the literature and to collect and classify relevant research in ``shoeboxes'', according to the different topics discussed in our survey.
For instance, concerning the visualization techniques on LHRDs, we built distinct shoeboxes for specific types of data as shown in Figure~\ref{fig:papyrusShoeboxes}-b.
In addition to the formally collected literature, the co-authors also contributed further references based on their individual scientific background.

\subsection{Software \& Open Data}

Throughout our process, we used several tools, Microsoft Excel to create a spreadsheet with the papers. We developed macros in VBA to clean up the corpus and the Fuzzy lookup add-in to deduplicate conference and journal names. This corpus was loaded and processed afterwards in Papyrus, a visual text analytics tool developed internally.

Our complete dataset and tagging system are accessible online at \url{https://viswallsurvey.list.lu}. We also provide the final spreadsheet with the 701 papers underlying this survey in supplementary material and online at \url{https://dx.doi.org/10.21227/1aqm-xr17}. We invite other researchers to contribute to this corpus and join the discussion about wall-sized displays by contacting us.

\section{Additional tables}
\label{app:tables}
\begin{table}[t!]
    \caption{LHRD visualizations for different data types and how they utilize the increased display space.}
    \mycfs{6}
    
	\setlength\tabcolsep{0.10cm}
	\def\arraystretch{0.99}
	\label{tab:visualization_tab}
	\begin{tabularx}{\linewidth}{@{} Xl bbbbbb aaaa}
	\hline\\[-6pt]
        \textbf{First author} & \textbf{Year} & 
        \rot{\textbf{Geographical}} & 
        \rot{\textbf{Spatial}} & 
        \rot{\textbf{Temporal}} & 
        \rot{\textbf{Multidimensional}} & 
        \rot{\textbf{Networks and trees}} & 
        \rot{\textbf{Text and documents}} &
        
        \rot{\textbf{Single view}} &
        \rot{\textbf{Small multiples}} &
        \rot{\textbf{Multiple views}} &
        \rot{\textbf{Distributed views}} \\
	\hline\\[-6pt]

    Ball~\cite{ball2005evaluating}           & 2005 &\tB&\tb&\tb&\tb&\tb&\tb & \tB&\tb&\tb&\tb\\
    Prouzeau~\cite{prouzeau2018awareness}        & 2018 &\tB&\tb&\tb&\tb&\tb&\tb & \tB&\tb&\tb&\tb\\
    Ball~\cite{ball2007move}                 & 2007 &\tB&\tb&\tb&\tb&\tb&\tb & \tB&\tb&\tb&\tb\\
    Jakobsen~\cite{jakobsen2013information}      & 2013 &\tB&\tb&\tb&\tB&\tb&\tb & \tB&\tb&\tB&\tb\\
    Shoemaker~\cite{shoemaker2010body}            & 2010 &\tB&\tb&\tb&\tb&\tb&\tb & \tB&\tb&\tb&\tb\\
    Ball~\cite{ball2008effects}              & 2008 &\tB&\tb&\tb&\tb&\tb&\tb & \tB&\tb&\tb&\tb\\
    Nancel~\cite{nancel2011mid}                & 2011 &\tB&\tb&\tb&\tb&\tb&\tb & \tB&\tb&\tb&\tb\\
    Onorati~\cite{onorati2015walltweet}         & 2015 &\tB&\tb&\tb&\tb&\tb&\tB & \tB&\tb&\tB&\tb\\
    Shoemaker~\cite{shoemaker2010whole}           & 2010 &\tB&\tb&\tb&\tb&\tb&\tb & \tB&\tb&\tb&\tb\\
    Ball~\cite{ball2007realizing}            & 2007 &\tB&\tb&\tb&\tb&\tb&\tb & \tB&\tb&\tb&\tb\\
    Ruddle~\cite{ruddle2015performance}        & 2015 &\tB&\tb&\tb&\tb&\tb&\tb & \tB&\tb&\tb&\tb\\
    Booker~\cite{booker2007high}               & 2007 &\tB&\tb&\tb&\tb&\tb&\tb & \tB&\tb&\tb&\tb\\
    Ronne~\cite{ronne2011sizing}              & 2011 &\tB&\tb&\tb&\tb&\tb&\tb & \tB&\tb&\tb&\tb\\
    Shupp~\cite{shupp2006evaluation}          & 2006 &\tB&\tb&\tb&\tb&\tb&\tb & \tB&\tb&\tb&\tb\\
    Dostal~\cite{dostal2014spidereyes}         & 2014 &\tB&\tb&\tb&\tb&\tb&\tb & \tB&\tb&\tb&\tb\\
    Isenberg~\cite{isenberg2013hybrid}           & 2013 &\tB&\tb&\tB&\tb&\tB&\tb & \tB&\tB&\tb&\tb\\
    Chung~\cite{chung2013developing}          & 2013 &\tB&\tb&\tb&\tb&\tb&\tb & \tB&\tb&\tb&\tb\\
    
    Chokshi~\cite{chokshi2014eplan}             & 2014 &\tB&\tb&\tb&\tb&\tb&\tB & \tb&\tB&\tB&\tB\\
    Reda~\cite{reda2015effects}              & 2015 &\tB&\tB&\tB&\tB&\tb&\tb & \tB&\tB&\tB&\tb\\
    
    Horak~\cite{horak2018david}               & 2018 &\tB&\tb&\tb&\tB&\tb&\tb & \tb&\tb&\tB&\tB\\
    Prouzeau~\cite{prouzeau2016towards}          & 2016 &\tB&\tb&\tb&\tb&\tb&\tb & \tb&\tb&\tB&\tb\\
    Ion~\cite{ion2013canyon}                & 2013 &\tB&\tb&\tb&\tb&\tb&\tb & \tb&\tb&\tB&\tb\\
    Kobayashi~\cite{kobayashi2018sage}            & 2018 &\tB&\tb&\tb&\tB&\tb&\tb & \tb&\tb&\tB&\tb\\
    Langner~\cite{langner2018multiple}          & 2018 &\tB&\tb&\tB&\tB&\tB&\tb & \tb&\tb&\tB&\tb\\
    Su~\cite{su_visually_2018}             & 2018 &\tB&\tb&\tb&\tb&\tb&\tB & \tb&\tB&\tB&\tb\\
    
    von Zadow~\cite{von2014sleed}                 & 2014 &\tB&\tb&\tb&\tb&\tb&\tb & \tb&\tb&\tb&\tB\\

    Treanor~\cite{treanor2009virtual}           & 2009 &\tb&\tB&\tb&\tb&\tb&\tb & \tB&\tb&\tb&\tb\\
    Goodyer~\cite{goodyer2009using}             & 2009 &\tb&\tB&\tb&\tb&\tb&\tb & \tB&\tb&\tb&\tb\\
    Ruddle~\cite{ruddle2016design}             & 2016 &\tb&\tB&\tb&\tb&\tb&\tb & \tB&\tb&\tb&\tb\\
    Reda~\cite{reda2013visualizing}          & 2013 &\tb&\tB&\tb&\tb&\tb&\tb & \tB&\tB&\tb&\tb\\
    Sollich~\cite{sollich2016exploring}         & 2016 &\tb&\tB&\tb&\tb&\tb&\tb & \tB&\tb&\tb&\tB\\
    Polys~\cite{polys2007effects}             & 2007 &\tb&\tB&\tb&\tb&\tb&\tb & \tB&\tb&\tb&\tb\\
    Hanula~\cite{hanula2015cavern}             & 2015 &\tb&\tB&\tb&\tb&\tb&\tb & \tB&\tb&\tb&\tb\\
                                        
    Liu~\cite{liu2014effects}               & 2014 &\tb&\tB&\tb&\tb&\tb&\tB & \tb&\tB&\tb&\tb\\
    Khan~\cite{khan2004remote}               & 2004 &\tb&\tB&\tb&\tb&\tb&\tb & \tb&\tB&\tb&\tb\\
    Jagodic~\cite{jagodic2011enabling}          & 2011 &\tb&\tB&\tb&\tb&\tb&\tB & \tb&\tB&\tb&\tB\\
                                        
    Pietriga~\cite{pietriga2016exploratory}      & 2016 &\tb&\tB&\tb&\tb&\tb&\tb & \tb&\tb&\tB&\tb\\
    Wigdor~\cite{wigdor2009wespace}            & 2009 &\tb&\tB&\tb&\tb&\tb&\tb & \tb&\tb&\tB&\tb\\
                                        
                                        
    Fernando~\cite{fernando_towards_2020}        & 2020 &\tb&\tb&\tB&\tb&\tb&\tB & \tB&\tb&\tb&\tb\\
                                        
                                        
    Badam~\cite{badam2016supporting}          & 2016 &\tb&\tb&\tB&\tb&\tb&\tb & \tb&\tb&\tB&\tb\\
    Thomas~\cite{thomas2017echo}               & 2017 &\tb&\tb&\tB&\tB&\tb&\tb & \tb&\tB&\tB&\tb\\
                                        
    Jansen~\cite{jansen2012tangible}           & 2012 &\tb&\tb&\tb&\tB&\tb&\tb & \tB&\tb&\tb&\tb\\
    Ruddle~\cite{ruddle2013leveraging}         & 2013 &\tb&\tb&\tb&\tB&\tb&\tb & \tB&\tb&\tb&\tb\\
    Mateescu~\cite{mateescu2015awall}            & 2015 &\tb&\tb&\tb&\tB&\tb&\tb & \tB&\tb&\tb&\tb\\
    Aurisano~\cite{aurisano2015bactogenie}       & 2015 &\tb&\tb&\tb&\tB&\tb&\tb & \tB&\tb&\tb&\tb\\
    Reibert~\cite{Reibert2020-PACM}             & 2020 &\tb&\tb&\tb&\tB&\tb&\tb & \tB&\tb&\tb&\tb\\
                                        
    Anslow~\cite{anslow2010user}               & 2010 &\tb&\tb&\tb&\tB&\tb&\tb & \tb&\tB&\tb&\tb\\
    Bezerianos~\cite{bezerianos2012perception}     & 2012 &\tb&\tb&\tb&\tB&\tb&\tb & \tb&\tB&\tb&\tb\\
    Yost~\cite{yost2007beyond}               & 2007 &\tb&\tb&\tb&\tB&\tb&\tb & \tb&\tB&\tb&\tb\\
    Chegini~\cite{chegini2017interaction}       & 2017 &\tb&\tb&\tb&\tB&\tb&\tb & \tb&\tB&\tB&\tb\\
                                        
    Tsandilas~\cite{tsandilas2015sketchsliders}   & 2015 &\tb&\tb&\tb&\tB&\tb&\tb & \tb&\tb&\tB&\tb\\
    Riehmann~\cite{Riehmann2020-CGF}             & 2020 &\tb&\tb&\tb&\tB&\tb&\tb & \tb&\tb&\tB&\tb\\
                                        
                                        
    Kister~\cite{kister2017grasp}              & 2017 &\tb&\tb&\tb&\tb&\tB&\tb & \tB&\tb&\tb&\tB\\
    Lehmann~\cite{lehmann2011physical}          & 2011 &\tb&\tb&\tb&\tb&\tB&\tb & \tB&\tb&\tb&\tb\\
    Kister~\cite{kister2015bodylenses}         & 2015 &\tb&\tb&\tb&\tb&\tB&\tb & \tB&\tb&\tb&\tb\\
    Prouzeau~\cite{prouzeau2016evaluating}       & 2016 &\tb&\tb&\tb&\tb&\tB&\tb & \tB&\tb&\tb&\tb\\
    von Zadow~\cite{von2017giant}                 & 2017 &\tb&\tb&\tb&\tb&\tB&\tb & \tB&\tb&\tb&\tb\\
    Isenberg~\cite{isenberg2009coconuttrix}      & 2009 &\tb&\tb&\tb&\tb&\tB&\tb & \tB&\tb&\tb&\tb\\
    Nolte~\cite{nolte2016collaborative}       & 2016 &\tb&\tb&\tb&\tb&\tB&\tb & \tB&\tb&\tb&\tB\\
    Chung~\cite{chung2014visporter}           & 2014 &\tb&\tb&\tb&\tb&\tB&\tB & \tB&\tb&\tB&\tB\\

                                        
                                        
    Bradel~\cite{bradel2013large}              & 2013 &\tb&\tb&\tb&\tb&\tB&\tB & \tb&\tB&\tb&\tB\\
                                        
    Iyer~\cite{iyer_text_2017}               & 2017 &\tb&\tb&\tb&\tb&\tb&\tB & \tB&\tb&\tb&\tb\\
    Nutsi~\cite{nutsi_readability_2016}       & 2016 &\tb&\tb&\tb&\tb&\tb&\tB & \tB&\tb&\tb&\tb\\
                                        
    Doshi~\cite{doshi2017stickyschedule}      & 2017 &\tb&\tb&\tb&\tb&\tb&\tB & \tb&\tB&\tb&\tb\\
    Jansen~\cite{jansen2019effects}            & 2019 &\tb&\tb&\tb&\tb&\tb&\tB & \tb&\tB&\tb&\tb\\
    Jakobsen~\cite{jakobsen2014up}               & 2014 &\tb&\tb&\tb&\tb&\tb&\tB & \tb&\tB&\tb&\tb\\
    Birnholtz~\cite{birnholtz2007exploratory}     & 2007 &\tb&\tb&\tb&\tb&\tb&\tB & \tb&\tB&\tb&\tb\\
    Kirshenbaum~\cite{kirshenbaum2019side}          & 2019 &\tb&\tb&\tb&\tb&\tb&\tB & \tb&\tB&\tb&\tb\\
    Lischke~\cite{lischke2015using}             & 2015 &\tb&\tb&\tb&\tb&\tb&\tB & \tb&\tB&\tb&\tb\\
    Jakobsen~\cite{jakobsen2012proximity}        & 2012 &\tb&\tb&\tb&\tb&\tb&\tB & \tb&\tB&\tb&\tb\\
    Andrews~\cite{andrews2010space}             & 2010 &\tb&\tb&\tb&\tb&\tb&\tB & \tb&\tB&\tb&\tb\\
    Endert~\cite{endert_semantic_2012-1}       & 2012 &\tb&\tb&\tb&\tb&\tb&\tB & \tb&\tB&\tb&\tb\\
                                        
    Geymayer~\cite{geymayer_how_2017}            & 2017 &\tb&\tb&\tb&\tb&\tb&\tB & \tb&\tb&\tB&\tb\\
    Fiaux~\cite{fiaux_bixplorer_2013}         & 2013 &\tb&\tb&\tb&\tb&\tb&\tB & \tb&\tb&\tB&\tb\\
    Andrews~\cite{andrews_analysts_2012}        & 2012 &\tb&\tb&\tb&\tb&\tb&\tB & \tb&\tb&\tB&\tb\\
    Kukimoto~\cite{kukimoto_open_2014}           & 2014 &\tb&\tb&\tb&\tb&\tb&\tB & \tb&\tb&\tB&\tB\\
    Lischke~\cite{lischke_towards_2017}         & 2017 &\tb&\tb&\tb&\tb&\tb&\tB & \tb&\tb&\tB&\tB\\
    \\[-6pt]
	\hline
\end{tabularx}
\end{table}


\begin{sidewaystable}[t!]
	\centering
	\mycfs{10.5}
    \caption{Interaction techniques }
	\setlength\tabcolsep{0.05cm}
	\def\arraystretch{0.99}
	\label{tab:interaction_tab}
	\begin{tabularx}{\linewidth}{@{} l l l @{}}
	    \hline 
	    \\[-6pt]
	    \textbf{\multirow[t]{3}{*}{On-surface interaction}}
	    & Touch and multi-touch
	    & \cite{jakobsen2014up,jakobsen2016negotiating,heidrich2011interacting,langner2018multiple,jakobsen2015should,prouzeau2016evaluating,Reibert2020-PACM,Riehmann2020-CGF}
	    \\
	    
	    & Pen input, pen and touch
	    & \cite{guimbretiere2001fluid,Walny2012-TVCG,ion2013canyon,lee2015sketchinsight}
	    \\

	    & Tangible input 
	    & \cite{Zadow2016miners,courtoux2021walltokens}
	    \\
	    \\[-6pt]
	    
	    \textbf{\multirow[t]{4}{*}{Distant interaction}} 
	    & Mouse
	    & \cite{johanson2002pointright,andrews2010space,birnholtz2007exploratory,jakobsen2016negotiating}
	    \\
	    
	    & Remote controller
	    &\cite{baudisch2006soap,davis2002lumipoint,konig2007position,jota2009comparative,jota2010comparison,jansen2012tangible,nancel2013high,zhang2017combining,langner2018multiple, horak2018david}\\

	    
	    & Mid-air gestures 
	    & \cite{malik2005interacting,vogel2005distant,nancel2011mid,yoo_dwell_2015,haque2015myopoint,liu2015gunslinger,jakobsen2015should,wittorf2016eliciting,matulic2018multiray}
	    \\

	    & Gaze 
	    & \cite{herholz2008libgaze,chuang2010measuring,stellmach2013still,lander2015gazeprojector}
	    \\
	    \\[-6pt]
	    
	    \textbf{\multirow[t]{3}{*}{Utilizing space}} 
	    & Physical navigation
	    & \cite{ball2007move,ball2007realizing,bi2009comparing,lischke2015using,endert2011visual,isenberg2013hybrid,Nacenta12FatFonts,liu2014effects,jakobsen2015moving,radle2013effect,jansen2019effects}
	    \\
	    
	    & Proxemic interaction
	    & \cite{jakobsen2013information,lehmann2011physical,dostal2014spidereyes,kister2015bodylenses,shoemaker2010whole,shoemaker2010body,prouzeau2018awareness}
	    \\

	    & Spatial interaction
	    & \cite{radle2013effect,langner2016content,kister2017grasp}
	    \\
	    \\[-6pt]
	    
	    \textbf{\multirow[t]{3}{*}{Multiple displays}} 
	    & Mobile devices 
	    & \cite{von2014sleed,von2015using,chapuis2014smarties,tsandilas2015sketchsliders,kister2017grasp,horak2018david}
	    \\
	    
	    & Head-mounted displays 
	    & \cite{nagao2016enabling,sun2019collaborative,Alsaiari2019-SMC,Reipschlaeger2020-TVCG}
	    \\

	    & Multiple devices
	    & \cite{chokshi2014eplan,prouzeau2018awareness,radloff2015supporting,Alsaiari2019-SMC}
	    \\
	    \\[-6pt]
	    \hline
	    
	\end{tabularx}
\end{sidewaystable}

\begin{sidewaystable*}[htbp]
	\caption{Overview of the seven evaluation scenarios by Lam et al.~\cite{lam2011empirical} and corresponding examples of concrete evaluation questions.}
	\mycfs{7}
	\setlength\tabcolsep{0.10cm}
	\def\arraystretch{0.99}
	\label{tab:evaluation_questions_examples}
	\begin{tabularx}{\linewidth}{l l l l}
		\hline
        \\[-6pt]
		\multicolumn{2}{l}{\textbf{Scenario}} & \textbf{Research questions} & \textbf{Citations} \\
        \\[-6pt]
		\hline     
        \\[-4pt]
		1 & UWP  & What is the everyday use of LHRDs? How do users process their work on LHRDs?                                                                         & ~\cite{ball2005analysis,bi2009comparing,bi2014walltop} \\
          &      & When and how could LHRDs facilitate and support the current workflow?                                                                                & \cite{rajabiyazdi2015understanding} \\
          &      & What are the use cases and benefits of LHRDs in a given situation?                                                                                   & \cite{chattopadhyay2018shared,ball2005evaluating,mateescu2015awall} \\
          &      & How do users organize and manage information on an LHRD?                                                                                             & \cite{bradel2013large,doshi2017stickyschedule} \\
          &      & What interaction and visualization would better fit with LHRDs? What are the challenges when combining an LHRD and other devices?                    & \cite{kister2017grasp} \\
          &      & How might LHRDs be a part of existing analysis tasks and practices?                                                                                  & \cite{knudsen2012exploratory} \\
          &      & What office work practices might benefit from LHRDs? & \cite{Knudsen2019-VAHC} \\
	   \\[-3pt]

        2 & VDAR & What are the user's data analysis rationale, thought process and findings when using an LHRD?                                                        & \cite{ruddle2013leveraging} \\
          &      & What is the effect of display size of LHRDs on the quantity and breadth of insights acquired by users during visual exploration?                     & \cite{reda2015effects,lischke2015using,shupp2009shaping,ruddle2015performance,ball2005analysis,rajabiyazdi2015understanding} \\
          &      & How do users adapt their sensemaking workflow to take advantage of an LHRD and its user interface?                                                   & \cite{reda2012scalable,reda2014expanding} \\
	   \\[-3pt]
        
        3 & CTV  & How effective is the displayed app for the task using an LHRD?                                                                                           & \cite{anslow2010user} \\
          &      & How users perceive and interact with interfaces and visualizations displayed on LHRD?                                                                & \cite{polys2007effects} \\
          &      & How perceptually scalable are data visualizations for large displays?                                                                                & \cite{yost2006perceptual} \\
	   \\[-3pt]
        
        4 & CDA  & How pairs collaborate on multi-touch LHRD?                                                                                                           & \cite{langner2018multiple,jakobsen2012proximity,jakobsen2014up} \\
          &      & What are the different collaborative position patterns of pairs working with LHRD and multiple mobile devices?                                       & \cite{altarawneh2015collaborative} \\
          &      & How group processes vary depending on the use of different modalities (e.g., direct interaction with touch or indirect distant interaction)? & \cite{jakobsen2016negotiating,hawkey2005proximity} \\
          &      & How groups behave when they are closely collaborating in front of an LHRD?                                                                              & \cite{Zadow2016miners,mayer2018pac} \\
          &      & How do the users collaborate and develop territories?                                                                                                & \cite{bradel2013large} \\
          &      & How users define interpersonal and onscreen workspaces?                                                                                              & \cite{wallace2016creating} \\
          &      & How collaborative interaction affects efficiency?                                                                                                    & \cite{liu2016shared} \\
	   \\[-3pt]
        
        5 & UP   & How display size and physical navigation affect task performance?                                                                                    & \cite{liu2014effects,jakobsen2013interactive,ronne2011sizing,shupp2009shaping,ball2007move,jakobsen2015moving,ball2008effects,ball2005effects,czerwinski2003toward} \\
          &      & How does an LHRD compare to a regular display for achieving a task?                                                                                   & \cite{ruddle2016design,treanor2009virtual} \\
          &      & What types of interaction zones users expected when interacting with displays?                                                                        & \cite{paul2018size} \\
          &      & How does visual encoding impact the effectiveness of large, high-resolution visualizations?                                                               & \cite{endert2011visual} \\
          &      & How does one interaction technique on LHRD compare to another by measuring the human performance?                                                    & \cite{nancel2011mid,jakobsen2015should,nancel2013high,jota2010comparison,jota2009comparative} \\ 
	   \\[-3pt]
        
        6 & UE   & Are the features useful for end-users? Are desired actions supported?                                                                            & \cite{thomas2017echo,chokshi2014eplan,ion2013canyon,anslow2010user,nolte2016collaborative,kister2017grasp,agarwal2019viswall} \\
          &      & What are the pain points users encounter during their process using the software on LHRD, and what improvements should be made?               & \cite{doshi2017stickyschedule,wigdor2009wespace,prouzeau2016towards} \\
	   \\[-3pt]
        
        7 & AP   & What is the performance of the algorithm loading an image on tiled LHRD?                                                                            & \cite{krishnaprasad2004juxtaview} \\
          &      & Which algorithm perform better for rendering (e.g., graph) visualization?                                                                           & \cite{mueller2006distributed,horak2018comparing,ren2017stardust,han2020virtual,martinez2020tuoris} \\ 
          &      & Which algorithm perform better for seamless frame synchronisation on tiled LHRD?                                                                     & \cite{nam2010multi} \\
	   \\[-4pt]
        \hline
	\end{tabularx}
\end{sidewaystable*}

\end{document}